\documentclass[english,showpacs,twocolumn,superscriptaddress,showpacs,showkeys]{revtex4-1}
\usepackage{graphicx}% Include figure files
\usepackage{chngcntr}
\usepackage[export]{adjustbox}
\usepackage{rotating}
\usepackage{dcolumn}% Align table columns on decimal point
\usepackage{bm}% bold math
\usepackage[usenames,dvipsnames]{color}
\usepackage{amsmath}
\usepackage{amssymb}
\usepackage{subcaption}
\usepackage[utf8]{inputenc}
\usepackage[colorlinks=true,linkcolor=blue,citecolor=green,urlcolor=magenta]{hyperref} %(For colored referencing)
\usepackage[nameinlink]{cleveref} %(For colored referencing)
\setcitestyle{super}
\graphicspath{{./Files/}}

\begin{document}

%=========================================================================================

\title{Effect of flow shear on the onset of dynamos}
%\vspace{8mm}
%\large
\author{Shishir Biswas} 
\email{shishirbeafriend@gmail.com}
\email{shishir.biswas@ipr.res.in}
\affiliation{Institute for Plasma Research, Bhat, Gandhinagar, Gujarat  382428, India}
\affiliation{Homi Bhabha National Institute, Training School Complex, Anushaktinagar, Mumbai 400094, India}

\author {Rajaraman Ganesh}
\email{ganesh@ipr.res.in}
\affiliation{Institute for Plasma Research, Bhat, Gandhinagar, Gujarat 382428, India}
\affiliation{Homi Bhabha National Institute, Training School Complex, Anushaktinagar, Mumbai 400094, India}
\date{\today}
%\author{Rupak Mukherjee} 
%\email{rupakmukherjee01@gmail.com}
%\affiliation{Princeton Plasma Physics Laboratory, Princeton, NJ - 08540, USA}
%
%\author {Abhijit Sen}
%\email{abhijit@ipr.res.in}
%\affiliation{Institute for Plasma Research, HBNI, Bhat, Gandhinagar, Gujarat  382428, India}
%\affiliation{Homi Bhabha National Institute, Training School Complex, Anushaktinagar, Mumbai 400094, India}

%\author {Naga Vijayalakshmi Vydyanathan}
%\email{nvydyanathan@nvidia.com}
%\author {Bharat k. Sharma}
%\email{bharatk@nvidia.com}
%\affiliation{NVIDIA, Bengaluru, Karnataka-560045, India}

\begin{abstract}
%Understanding astrophysical magnetic fields is an ongoing challenge for theoretical, experimental, and computational astrophysicists. High magnetic fields are thought to be generated by the dynamo mechanism in a wide range of astrophysical bodies, including planets, galaxies, the interstellar medium, accretion disks, and the Sun. Through the motion of conducting fluids, dynamos generate magnetic fields on multiple scales. Many numerical and experimental studies in the lab have been conducted to better comprehend these astrophysical scenarios. 
Understanding the origin and structure of mean magnetic fields in astrophysical conditions is a major challenge. Shear flows often coexist in such astrophysical conditions and the role of flow shear on dynamo mechanism  is only beginning to be investigated. Here, we present a direct numerical simulation (DNS) study of the effect of flow shear on dynamo instability for a variety of base flows with controllable mirror symmetry (i.e, fluid helicity). Our observations suggest that for helical base flow, the effect of shear is to suppress the small scale dynamo (SSD) action, i.e, shear helps the large scale magnetic field to manifest itself by suppressing SSD action. For non-helical base flows, flow shear has the opposite effect of amplifying the small-scale dynamo action. The magnetic energy growth rate ($\gamma$) for non-helical base flows are found to follow an algebraic nature of the form, $\gamma = - aS + bS^\frac{2}{3}$, where  $a, b > 0$ are real constants and $S$ is the shear flow strength and $\gamma$ is found to be independent of scale of flow shear. Studies with different shear profiles and shear scale lengths for non-helical base flows have been performed to test the universality of our finding.  
	
%Understanding of astrophysical magnetic fields is being a ever standing challenge for theoretical, experimental al well as computational astrophysicist. The Dynamo mechanism is
%believed to be the key mechanism behind the existence of this high magnetic fields in various astrophysical bodies, for example, planets, galaxies, interstellar medium, accretion disks, also in the Sun. Dynamos generate multi-scale magnetic fields by the motion of conducting fluids.  Various laboratory experiments, as well as numerical studies have been performed to understand these astrophysical scenarios in detail. Shear flows often coexist in astrophysical conditions and the role of flow shear on the onset of dynamo is only beginning to be investigated. Here we present a detail study of the impact of large scale shear flows on dynamo instability for a range of flows whose mirror symmetry can be controlled using direct numerical simulation (DNS). Our observation suggests that for helical flow the effect of shear is only to suppress the small scale dynamo (SSD) action, in the other words shear helps the large scale dynamo field to manifest it self. For non-helical flows it acts exactly opposite, i.e. shear enhances the small scale dynamo action with some significant scaling. To rest the universality, study using various shear profiles and small scale flows have been performed. 
\end{abstract} 

\maketitle

%=========================================================================================
\section{Introduction}
Predicting the generation of multi-scale magnetic fields, in many astrophysical bodies, has been a long-standing theoretical question in astrophysical plasmas. Different theories have been proposed to account for the origin of these multi-scale magnetic fields.  For example, invoking magnetic induction due to the motion of conducting fluids, \cite{Parker_APJ:1955, Parker:1979} suggested these multi-scale magnetic fields are generated via a hydromagnetic dynamo process and maintained against resistive losses.

Depending on the length scales involved, dynamos may be classified into two broad categories :
Small Scale or fluctuation Dynamo (SSD) and Large Scale or mean field Dynamo (LSD). Unlike SSD, for LSDs a lack of reflectional symmetry is widely believed to be a necessary condition \cite{Rincon:2019}. Depending on the time scales, dynamos may also be categorized as Fast dynamos (growth rate remain finite in the range $R_m \to \infty$) and Slow dynamos (magnetic diffusion plays a significant role) \cite{childress_STF:1995, Rincon:2019}. Fast dynamos are further classified into two sub-categories as, `quick' dynamo and `pedestrian' dynamo \cite{Tobias_Cattaneo_JFM:2008}. For a `quick' dynamo magnetic energy growth rate achieves its maximum value quickly as a function of magnetic Reynolds number $R_m$, where as for a `pedestrian' dynamo the growth rate very weakly depends on $R_m$ \cite{Tobias_Cattaneo_JFM:2008}. Depending on the feedback strength of the magnetic field on to the flow field, dynamos are regarded as linear or non-linear. For example, a linear dynamo is one in which the magnetic field dynamics does not “back react” with the velocity field and the velocity field is either given or it obeys the Navier-Stokes equation \cite{Rincon:2019}. A nonlinear dynamo or self-consistent dynamo is when the nonlinear effects start to change the flow (once the magnetic field is large enough) to stop further magnetic field growth , that is, the flow and B-field “back react” on each other, typically leading to nonlinear saturation \cite{Rincon:2019}.

SSDs may also be defined as systems which sustain B-field fluctuations at scales smaller than the forcing scale \cite{Batchelor_SSD:1950, Kazantsev_SSD:1968, Zel'dovich_SSD:1984, Pouquet_SSD_PRL:1981,Schekochihin_SSD:2004, Skoutnev_Squire_Bhattcharya:2021}. The fluctuating magnetic fields found in galaxies and clusters, as well as in the solar photosphere may be regarded as due to SSDs. Oftentimes, the generated magnetic fields are also observed to be correlated on scales larger than the driving scale, resulting in LSD action \cite{Rincon:2019, Skoutnev_Squire_Bhattcharya_MNRAS:2022}. For instance, the solar magnetic field possesses a large-scale dipole component which is mostly aligned with the Sun's rotation axis and a wave of magnetic activity that traverses from mid-latitudes to the equator on an 11-year time scale is clearly visible in the solar butterfly diagram \cite{Stix_Sun_Butterfly:2004}. Large-scale dynamo activity can also be explained by the well-known $\alpha$ effect \cite{Moffatt:1978,Parker:1979, Rincon:2019}, provided the system has some mirror-symmetry breaking (i.e, when kinetic or fluid helicity is non-zero). 

Not only by the nature of turbulence, but, dynamos are also affected by factors such as density lamination \textcolor{black}{(for example, density variation along the direction of gravity)}, rotation, kinetic helicity (mirror symmetry breaking), and flow shear. Out of these factors, flow shear is ubiquitous in astrophysical systems - appearing in \textcolor{black}{interstellar medium}, galaxies, accretion disks, and in liquid-metal laboratory dynamo \cite{Monchaux_Dynamo_exp_PRL:2007} experiments. The paradigm of investigation of the exponential growth of magnetic field caused by the interaction of small-scale velocity fluctuations and a large-scale velocity shear; is commonly referred to as the ``shear dynamo problem''. For example, presence of a large-scale velocity shear, in association with turbulent rotating convection \textcolor{black}{(turbulent convective motion under the influence of rotation)}, is seen to actually increase the dynamo growth rate at larger scales \cite{Kapyl_Korpi_Brandenburg_AA:2008, Hughes_Proctor_PRL:2009, Kapyl_Korpi_Brandenburg_MNRAS:2010, Hughes_Proctor:2013}.  Furthermore, it is also found that a highly helical flow pattern may only result in a SSD action when the rotational convention is sufficiently strong \cite{Cattaneo_Hughes_JFM:2006}.

 For conditions where rotational effects are negligible, an integro-differential equation has been proposed based on a quasi-linear model to address the limit of weak convective flow \cite{Sridhar_Subramanian_PRE:2009a, Sridhar_Subramanian_PRE:2009b}. In order to further investigate the shear dynamo problem, several other analytical frameworks have been reported \cite{Sridhar_Singh:2010, Sridhar_Singh:2014, Singh_Sridhar:2014}. Along with these analytical attempts, it has been reported, based on direct numerical simulation that a driven \textcolor{black}{small}-scale, purely non-helical turbulence enhances the exponential growth of large scale magnetic energy in the presence of non-rotating linear shear flows \cite{Brandenburg_APJ:2008, Yousef_PRL:2008, Yousef:2008b}. For example, it is found that \cite{Yousef_PRL:2008, Yousef:2008b}, the LSD growth rate scales linearly with the $S$ (where $S$ is the shear flow strength). On the other hand, using a kinematic dynamo model, it has been shown that, unlike a linear relationship, the dynamo growth rate scales as $S^\frac{2}{3}$ \cite{Proctor_JFM:2012}.

It is clear from the preceding discussion that theoretical and computational efforts are being put to understand the origin of large-scale dynamo action. Numerical studies have also been carried out  on the shear dynamo action for large scale velocity shear and helical forced turbulence \cite{Kapyla_APJ:2009} and provide an effective explanation for large-scale dynamo action using a propagating wave-like dynamo solution \cite{Kapyla_APJ:2009}. The primary difficulty lies in controlling the fluctuations on a small scale, as the small scale magnetic fields are regarded to be harmful to the dynamo action on a larger scale \cite{Bhattcharya_PRL:2015}. In the presence of large-scale velocity shear and non-helical flows, recent work provides an evidence of large-scale magnetic field generation from small-scale dynamos \cite{Bhattcharya_PRL:2015, Bhattcharya_APJ:2015}. This intriguing numerical observation has been explained using the concept of ``magnetic shear current effect'' \cite{Rogachevskii_Shear_Current:2003, Rogachevskii_Shear_Current:2004, Bhattcharya_PRL:2015, Bhattcharya_APJ:2015}. The generation of large-scale magnetic fields is the primary focus of all of these studies.

An alternate school of thought for LSD is to decrease the efficiency of small scales rather than trying to increase the activity of large-scale dynamos  \cite{Tobias_Nature:2013, Cattaneo_APJ:2014, Nigro_MNRAS:2017}. Kinematic dynamo model has been used to examine shear dynamo activity with superimposed large-scale shear flow and small-scale helical base flow \cite{Tobias_Nature:2013, Cattaneo_APJ:2014}. For the numerical experiment, a well-known time-dependent 2.5-dimensional GP flow \citep{Galloway_Proctor:1992} has been considered. The presence of symmetry along one spatial dimension in GP flow \cite{Galloway_Proctor:1992} allows one to effectively transform the three-dimensional (3D) kinematic dynamo problem into a two-dimensional (2D) one. The results of numerical simulation led to the conclusion that the interaction between large-scale shear flow and small-scale helical flow does not boost the induction process. Instead, it slows down the small-scale dynamo growth rate, which in turn makes it possible for the large-scale dynamo action to become apparent \cite{Tobias_Nature:2013, Cattaneo_APJ:2014, Nigro_MNRAS:2017}. This idea is sometimes referred to as the ``suppression principle''. In addition, propagating dynamo waves \cite{Parker_APJ:1955} have been observed, which is a hallmark of large-scale dynamo activity \cite{Tobias_Nature:2013, Cattaneo_APJ:2014, Nigro_MNRAS:2017}. By taking magnetic feed back into consideration (non-linear dynamo action) this issue has been revisited \cite{Pongkitiwanichakul_APJ:2016, Teed_MNRAS:2016,Teed_MNRAS:2017}.

Recently, the shear-dynamo activity in the non-helical limit has been investigated numerically using both the kinematic and self-consistent (with magnetic feed back) dynamo models \cite{Singh_APJL:2017}.  In addition to a linear shear, the model also incorporates a random non-helical white-noise as a body force. This model has also been used to explain why existence of large-scale velocity shear is a favourable condition for small-scale dynamos. The turbulence caused by flow shear provides an explanation for the enhancement of small-scale dynamo \cite{Singh_APJL:2017}.

In the context of shear dynamo problem, the effect of flow shear on non-helical base flow is studies sparsely. In this present work, we have investigated the shear dynamo action using a kinematic dynamo model. In our model, the velocity field is not simulated using the Navier-Stokes equation; instead, it is given and remains unchanged throughout the simulation. As a flow drive for our simulation, we have considered a recently reported three-dimensional Yoshida-Morrison flow (or YM flow \cite{EPI2D:2017} in short).  One of the interesting features of YM flow is that, its mirror symmetry (kinetic helicity) can be controlled by varying the magnitude of certain flow parameter \cite{Biswas_arxiv:2022}. In the maximal helicity limit, YM flow resembles the well-known Arnold-Beltrami-Childress (ABC) flow, while in the non-helical limit, it is known as EPI2D flow \cite{EPI2D:2017}. The ABC flow, but not the EPI2D flow, is a well-known candidate for fast dynamo action. \textcolor{black}{Here, we investigate the effect of flow shear on dynamo instability. For a helical base flow, such as ABC flow, the inductive process is known to result in an exponential increase in magnetic energy in the absence of shear flows.} Our spectral analysis supports the notion that this dynamo operates on a relatively small scales. \textcolor{black}{We have also identified that, for this helical base flow (ABC flow), the presence of flow shear effectively suppresses small-scale dynamo activity over a broad range of the magnetic Reynolds number $R_m$.} 
Several authors \cite{Tobias_Nature:2013, Cattaneo_APJ:2014, Nigro_MNRAS:2017} have reported this suppression mechanism using a quasi-2D helical base flow (GP flow) \cite{Galloway_Proctor:1992} with a large scale shear.  Here, we have observed similar suppression activity using a full 3D helical ABC flow.

 The above said picture changes dramatically when EPI2D flow is considered as the base flow. \textcolor{black}{We find that the EPI2D flow is unable to induce exponential amplification of magnetic energy in the absence of shear flows. Interestingly, when the shear flow is considered, the small scale EPI2D flow is found to  generate exponentially growing magnetic energy with time. In other words, our numerical analysis suggests that, in the presence of shear flow, an otherwise non-dynamo producing non-helical base flow (EPI2D flow) can effectively generate fast dynamo activity.} We also observe, through numerical simulation, that the strength of shear flows has a significant impact on the amount of small-scale dynamo activity. We have obtained a generalized algebraic (combination of linear and non-linear) scaling, for the growth rate of magnetic energy with shear flow strength $S$ as $\gamma = - aS + bS^\frac{2}{3}$, where $a$ and $b$ are positive real constants.  It is observed that, our numerical finding of depending of $\gamma$ on $S$ is in agreement with several earlier analytical works \cite{Kolokolov:2011, Proctor_JFM:2012} while generalizing the same. The robustness of our numerical finding is tested using shear flows with varying shear length scales, in addition, for a number of different small-scale base flows. \textcolor{black}{Accretion discs, galaxies, jets, stellar convective zones, and so on all include hydrodynamic flows that are characterized by significant flow shear, suggesting that the dynamo mechanisms under consideration may play a key role in the creation of magnetic fields in these astrophysical scenarios.}

The organization of the paper is as follows. In Sec. II we present about the dynamic equations. Our numerical solver and simulation details are described in Sec. III. The initial conditions, parameter details are shown in Sec. IV. Section V is dedicated to the simulation results on induction dynamo action that we obtained from our code and finally the summary and conclusions are listed in Sec. VI.

%=========================================================================================

%=========================================================================================
\section{Governing Equations}\label{Equations}
The governing equations to study \textcolor{black}{kinematic fast} dynamo action for the single fluid MHD plasma are as follows,
\begin{eqnarray}
	%&& \label{density} \frac{\partial \rho}{\partial t} + \vec{\nabla} \cdot \left(\rho \vec{u}\right) = 0\\
	%&& \frac{\partial (\rho \vec{u})}{\partial t} + \vec{\nabla} \cdot \left[ \rho \vec{u} \otimes \vec{u} + \left(P + \frac{B^2}{2}\right){\bf{I}} - \vec{B}\otimes\vec{B} \right]\nonumber \\
	%&& \label{velocity} ~~~~~~~~~ = \mu \nabla^2 \vec{u}\\
	%&& P = C_s^2 \rho \\
	&& \label{Bfield} \frac{\partial \vec{B}}{\partial t} + \vec{\nabla} \cdot \left( \vec{u} \otimes \vec{B} - \vec{B} \otimes \vec{u}\right) = \textcolor{black}{\frac{1}{R_m}} \nabla^2 \vec{B}\\
	&& \label{div B} \vec{\nabla} \cdot \vec{B} = 0
\end{eqnarray}
\textcolor{black}{where, $\vec{u}$, $\vec{B}$ and $R_m$ represent the velocity, magnetic fields and magnetic Reynolds number respectively. \textcolor{black}{The magnetic Reynolds number ($R_m$) is defined as, $R_m = \frac{u_0 L}{\eta}$, where $\eta$ is magnetic diffusivity and $u_0$ is a typical velocity scale. Time is normalized to Alfven times \textcolor{black}{(i.e, time taken for an Alfven wave to traverse the simulation domain)} and length to a typical characteristic length scale $L$ \textcolor{black}{(here it is the length of simulation domain)}.}  The symbol ``$\otimes$'' represents the dyadic between two vector quantities.}
\label{equations}

For solving the above set of equations at high grid resolution, we have developed a suite of GPU codes namely GMHD3D, which is briefly described in the following Section.

%=========================================================================================
\section{\textbf{Simulation Details: \textit{GMHD3D} Solver}}
In this Section, we discuss the details of the numerical solver  along with the  benchmarking of the solver carried out by us.
In order to study the plasma dynamics governed by MHD equations described above, we have recently upgraded an already existing well bench-marked single GPU MHD solver \cite{rupak_thesis:2019}, developed in house at Institute For Plasma Research to multi-node, multi-card (multi-GPU) architecture for better performance \cite{GTC}. This newly upgraded GPU based magnetohydrodynamic solver (\textit{GMHD3D}) is now capable of handling very large grid sizes. \textit{GMHD3D} is a multi-node, multi-card, three dimensional (3D), weakly compressible, pseudo-spectral, visco-resistive solver \cite{GTC}. This suite (GMHD3D) includes both 2-dimensional and 3-dimensional HydroDynamic (HD) and MagnetoHydrodynamic (MHD) solvers. It uses pseudo-spectral technique to simulate the dynamics of 3D magnetohydrodynamic plasma in a cartesian box with periodic boundary condition. By this technique one  calculates the spatial derivative to evaluate non-linear term in governing equations with a standard $\frac{2}{3}$ de-aliasing rule \cite{dealiasing:1971}. OpenACC FFT library (AccFFT library \cite{Accfftw:2016}) is used to perform Fourier transform and Adams-bashforth time solver, for time integration. For 3D iso-surface visualization, an open source Python based data converter to VTK (Visualization Tool kit) by ``PyEVTK'' \cite{VTK} is developed, which converts ASCII data to VTK binary format. After dumping the state data files to VTK, an open source visualization softwares, VisIt 3.1.2 \cite{visit} and Paraview \cite{paraview} is used to visualize the data.
For this present work, the new solver's accuracy with the single GPU solver has been cross-checked and it is verified that the results match upto machine precision. Further, \textcolor{black}{several other} benchmarking studies have been performed such as, the 3D kinematic dynamo effect \cite{Frish_Dynamo:1986, Dorch:2000, Archontis:2003, Bouya:2013}, have been reproduced with ABC flow at grid resolution $64^3$. \textcolor{black}{Details of these are presented in Appendix \ref{Appen A}.} \textcolor{black}{As will be discussed in the coming Section,  numerical simulations reported here are performed in $256^3$ grid size.}

As discussed in the Introduction, to study the kinematic fast dynamo action, an accurate selection of ``drive'' velocity field is crucial, which we discuss in the Section to follow.   
%=========================================================================================
\section{Initial Condition}

 Recently Yoshida and Morrison \cite{EPI2D:2017} (YM) proposed a new intermediate class of flow, which may be regarded as a topological bridge between quasi-2D  and 3D flow classes.  The flow is formulated as follows:
 \begin{equation} \label{base flow}
 	\vec{u}_b = u_0 \alpha    \vec{u}_+ + u_0 \beta    \vec{u}_-
 \end{equation}
 with
 \begin{align} 
 	\vec{u}_+ &= \begin{bmatrix} 
 		B sin(k_0y) - C cos(k_0z) \\
 		0\\
 		A sin(k_0x) \\
 	\end{bmatrix}
 \end{align}
 and 
 \begin{align} 
 	\vec{u}_- &= \begin{bmatrix} 
 		0\\
 		C sin(k_0z) - A cos(k_0x)\\
 		-B cos(k_0y) \\
 	\end{bmatrix}
 \end{align}
 so that,
 \begin{equation}\label{Yoshida_flow}
 	\begin{aligned}
 		u_x &= \alpha u_0 [ B \sin(k_0y) - C \cos(k_0z) ]\\
 		u_y &= \beta u_0 [ C \sin(k_0z) - A \cos(k_0x) ]\\
 		u_z &= u_0 [ \alpha A \sin(k_0x) - \beta B \cos(k_0y) ]
 	\end{aligned}
 \end{equation}
 where \textcolor{black}{$k_0$}, $\alpha, \beta$, A, B and C are  arbitrary real constants. \textcolor{black}{For the present study, we consider the value of $u_0, \alpha$, A, B and C to be unity.} In the present work, we consider Eq. \ref{base flow} as our base flow $\vec {u}_b$. The variation of $\beta$ value in YM flow leads to new classes of base flows.
 
 %\subsection{EPI-2D flow from YM flow ($\alpha = 1$, $\beta = 0$):} 
 For example, for \textcolor{black}{$\beta = 0$}, Yoshida et al. \cite{EPI2D:2017} classify this flow as EPI-2D flow which is given by :
 \begin{equation}\label{EPI2D}
 	\textcolor{black}{
 		\begin{aligned}
 			u_x &=  [ \sin(k_0y) - \cos(k_0z) ]\\
 			u_y &= 0\\
 			u_z &=  [ \sin(k_0x)]
 		\end{aligned}
 	}
 \end{equation}
 \textcolor{black}{This flow (i.e, Eq. \ref{EPI2D}) is dependent on all the 3 spatial coordinates (i.e, $x, y, z$), whereas only two flow components are nonzero. Thus EPI-2D flow is quasi-2D in nature.}  
 
 As can be expected, for $\beta = 1$ Eq. \ref{Yoshida_flow} becomes the well known Arnold–Beltrami–Childress flow [ABC] like flow, 
 \begin{equation}\label{ABC_like}
 	\textcolor{black}{
 		\begin{aligned}
 			u_x &=  [ \sin(k_0y) -  \cos(k_0z) ]\\
 			u_y &=  [  \sin(k_0z) -  \cos(k_0x) ]\\
 			u_z &=  [  \sin(k_0x) -  \cos(k_0y) ]
 		\end{aligned}
 	}
 \end{equation}
 As $\beta$ is varied from $0$ to $1.0$, a whole set of intermediate class of flows emerge, such that a normalized fluid helicity is exactly $0.0$ for $\beta = 0$ and is maximum for $\beta =1.0$ (i.e, ABC-like flows) \cite{Biswas_arxiv:2022}. The variation of $\beta$ value clearly leads to two distinguishable class viz helical ($\beta > 0$) and non-helical ($\beta=0$) class of base flows \cite{Biswas_arxiv:2022}.  
 
\begin{figure*}
	\centering
	\begin{subfigure}{0.49\textwidth}
		\centering
		%\hfill
		\includegraphics[scale=0.08480]{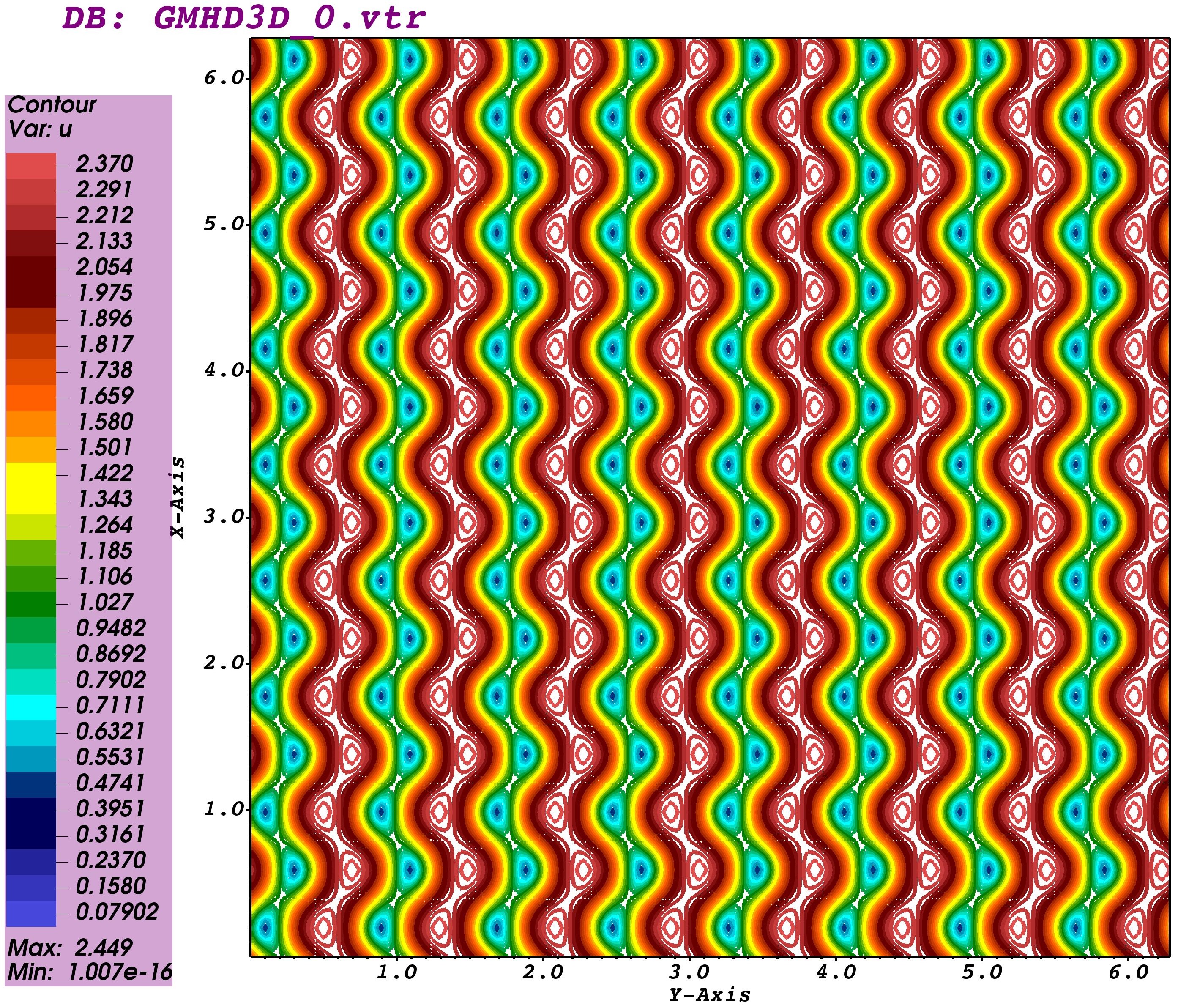}
		\caption{}
		%	\label{initial ABC flow}
	\end{subfigure}
	\begin{subfigure}{0.49\textwidth}
		\centering
		\includegraphics[scale=0.08480]{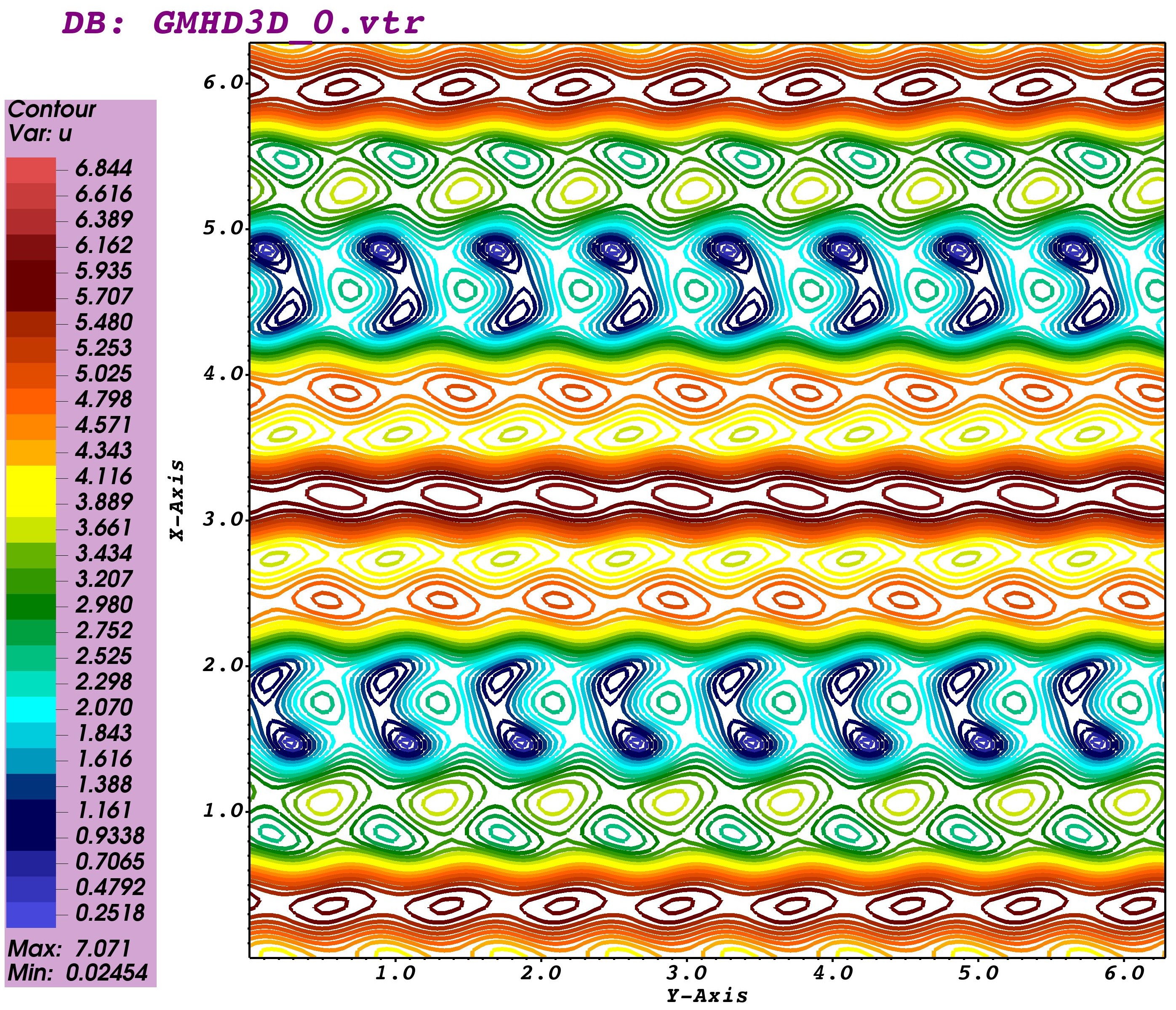}
		\caption{}
	%	\label{initial ABC flow with cos shear}
	\end{subfigure}
	\begin{subfigure}{0.49\textwidth}
		\centering
		\includegraphics[scale=0.08480]{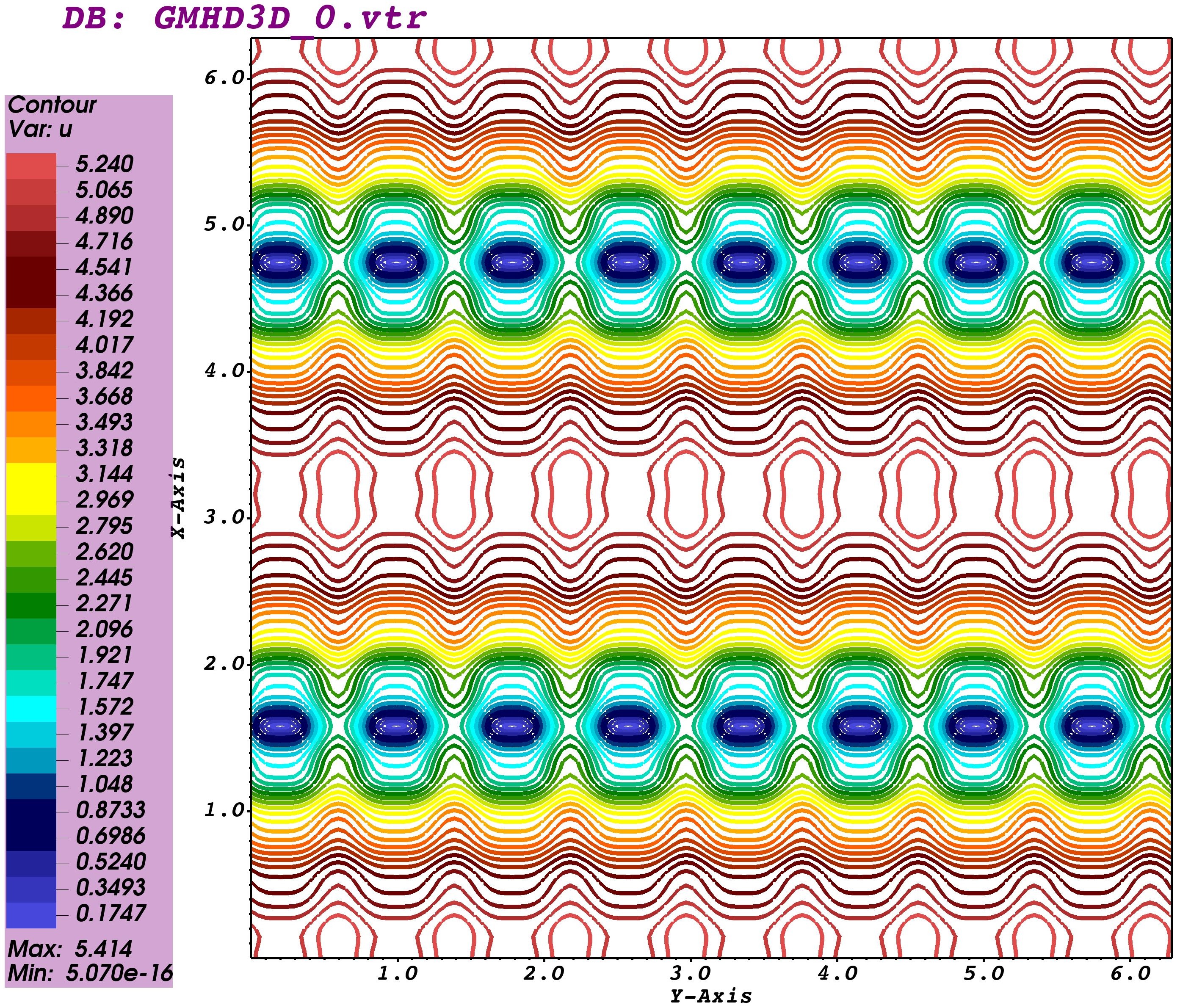}
		\caption{}
	%	\label{initial EPI2D flow with cos shear}
	\end{subfigure}
	\begin{subfigure}{0.49\textwidth}
		\centering
		\includegraphics[scale=0.08480]{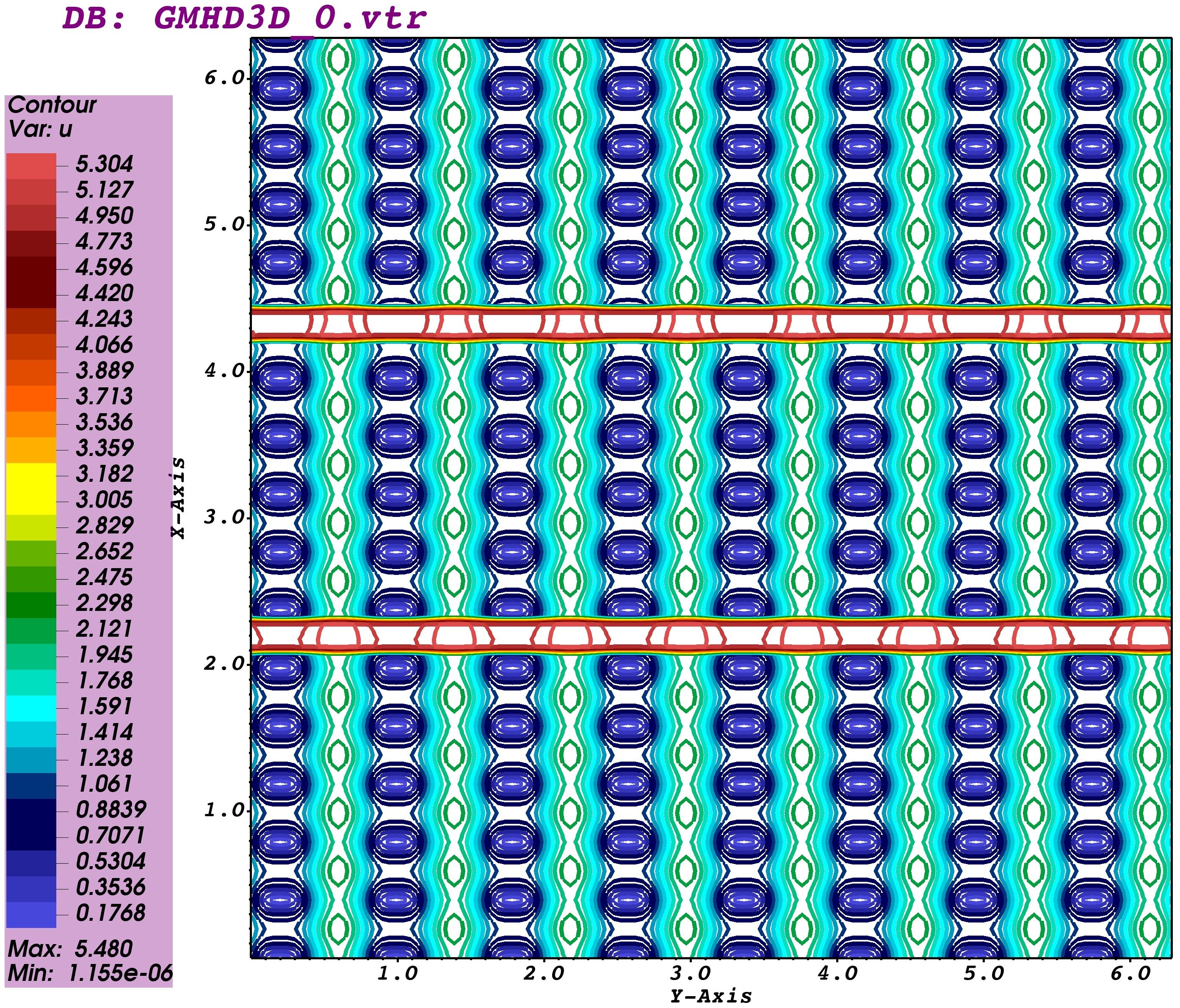}
		\caption{}
	%	\label{initial EPI2D flow with strip shear}
	\end{subfigure}
	\caption{Initial velocity ($u = \sqrt{u_x^2 + u_y^2 + u_z^2}$) profile in $X-Y$ plane for (a) small scale ($k_0 = 8$) helical ABC flow (b) superposition of small scale ($k_0 = 8$) helical ABC flow and large scale ($k_s = 1.0$) periodic shear (c) superposition of small scale ($k_0 = 8$) non-helical EPI2D flow and large scale ($k_s = 1$) periodic shear (d) superposition of small scale ($k_0 = 8$) non-helical EPI2D flow and broken jet (i.e $k_s \to \infty$) flow shear \cite{Drazin:1961, Thess_Strip:1994, Shishir_POF:2022} profile.}
	\label{initial ABC flow}
\end{figure*}

 In the following, we begin our investigation by focusing on the most well-known type of helical flow such as ABC-like base flow with $k_0 = 8$ and $\beta = 1$ (See Fig. \ref{initial ABC flow}a). In order to investigate the role of shear flows in the context of dynamo action, we introduce a periodic, large-scale shear flow (Eq. \ref{cos shear profile}) (See Fig. \ref{initial ABC flow}b) of the form 
 
 %As discussed recently, we first consider the maximum helical well known small scale ABC flow (See Fig. \ref{initial ABC flow}) for studying dynamo activity. To examine the effect of shear flows in the context of dynamo action we add a steady large-scale shear flow of the form
 \begin{equation}\label{cos shear profile}
 	\vec{u}_s = (0, S\cos (k_s x), 0)
 \end{equation}
 where $S$ is shear flow strength and $k_s$ is the length scale of shear flow such that $k_s < k_0$. Hence we refer to the base flow i.e, the $\beta = 1$ ABC flow with mode number $k_0$ as small-scale flow and the flow with $k_s<k_0$ as large scale shear flow. %along with the above discussed small scale ABC flow (See Fig. \ref{initial ABC flow with cos shear}).
 
 A recent work \cite{Biswas_arxiv:2022} has shown that a purely non-helical quasi-2-dimensional EPI2D flow alone is incapable of producing fast dynamo action due to insufficient stretching capability. \textcolor{black}{In the context of dynamo activity, it is interesting to examine the effect of large-scale shear flows on non-helical base flows i.e, EPI2D flow. Hence, we consider a small scale ($k_0 = 8$) EPI2D flow along with a periodic large scale ($k_s = 1$) shear flow (Eq. \ref{cos shear profile}) as our starting point (See Fig. \ref{initial ABC flow}c) to explore the shear dynamo action. In order to examine the robustness of our numerical findings, we have also considered a broken jet \textcolor{black}{(two jets with opposed directions i.e., broken jets whose width is $\frac{\pi}{16}$ in a system of length $2\pi$, placed alternately one after the other.)} flow shear profile \cite{Drazin:1961, Thess_Strip:1994, Shishir_POF:2022} instead of a periodic shear profile ($k_s \to \infty$) (See Fig. \ref{initial ABC flow}d) to investigate the shear dynamo activity. Few additional studies with smaller scales ($k_0 = 16$) base flow have been performed [Details of which are added in Supplementary information].}

An initial value problem involving the induction equation for $\vec{B}$ is solved for the prescribed flows $\vec{u} = \vec{u}_b + \vec{u}_s$ (See Eq. \ref{base flow} \& \ref{cos shear profile}). We have considered a random perturbation as a seed initial magnetic field for our numerical experiments. \textcolor{black}{We have also performed numerical experiments with a periodic initial magnetic field \cite{childress_STF:1995} and a uniform magnetic field, and we find that the characteristics of the dynamo are largely insensitive to the initial conditions in both cases.} For the rest of the discussion, we present results from random perturbations as initial magnetic field.\\

 \subsection{Parameter Details}
 We evolve the set of equations discussed in Section \ref{Equations}, for class of YM flow profile, in a triply periodic box of length $L_x = L_y = L_z = 2\pi$ with time steeping $(dt) = 10^{-4}$ and grid resolution $256^3$. We have also conducted grid size and time step size scaling studies (not shown) and find that values indicated above are adequate. With these initial conditions and parameter spaces we present our numerical simulation results.\\

%=========================================================================================
\section{Simulation Results}
%Kinematic dynamo action has been extensively studied with ABC flow \cite{Frish_Dynamo:1986, Dorch:2000, Archontis:2003, Bouya:2013, Biswas_arxiv:2022}. 
The helical nature, chaotic property, and stretching ability of the ABC-like flow are known to be the primary causes for the generation of dynamo action \cite{Frish_Dynamo:1986, Dorch:2000, Archontis:2003, Bouya:2013, Biswas_arxiv:2022}.

\begin{figure*}
	%	\centering
	\begin{subfigure}{0.49\textwidth}
		\centering
		%\hfill
		\includegraphics[scale=0.6]{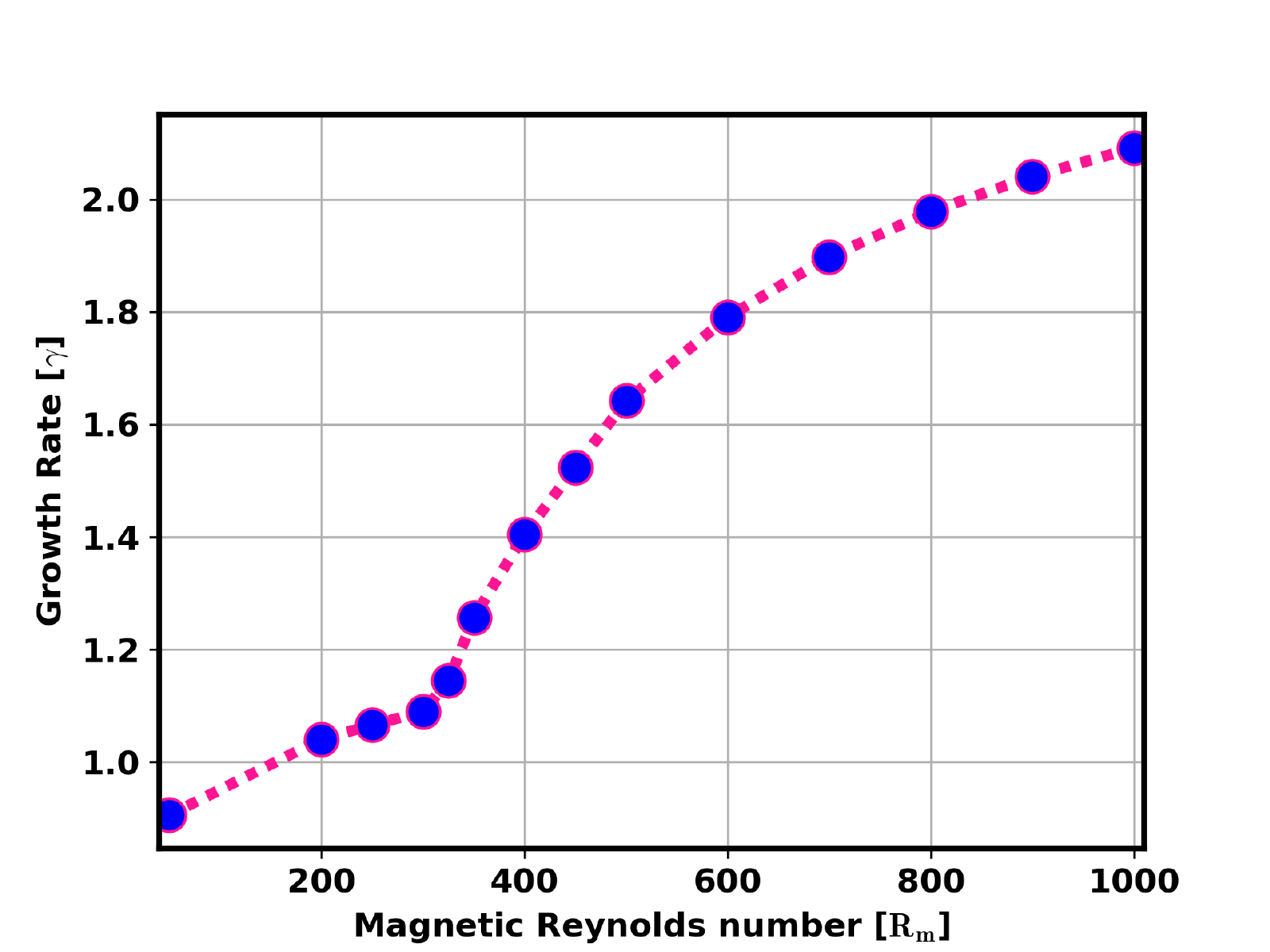}
		\caption{}
	%	\label{ABC flow growth rate}
	\end{subfigure}
	\begin{subfigure}{0.49\textwidth}
		\centering
		\includegraphics[scale=0.07]{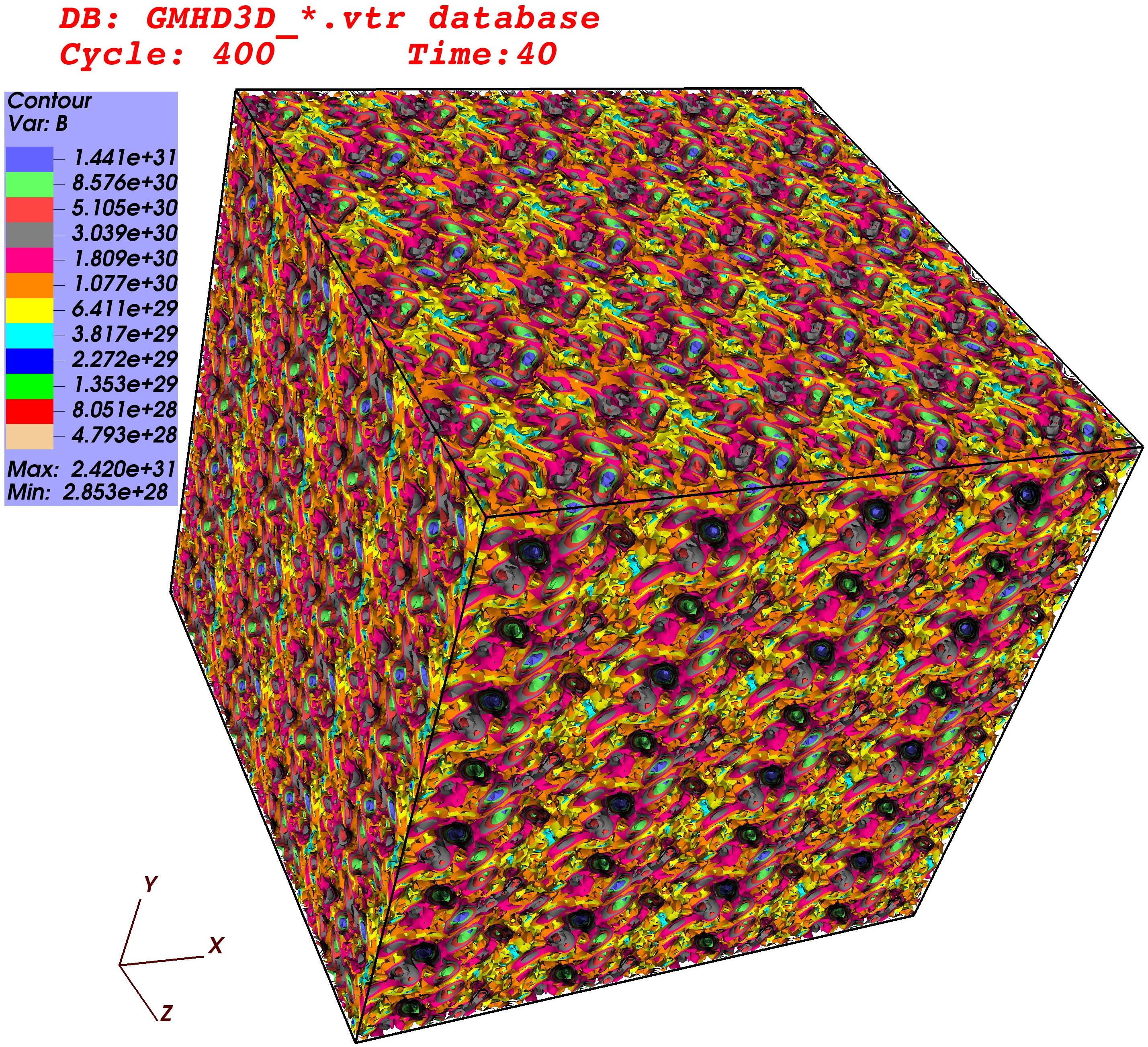}
		\caption{}
	%	\label{ABC flow isoB}
	\end{subfigure}
	\begin{subfigure}{0.49\textwidth}
		\centering
		\includegraphics[scale=0.6]{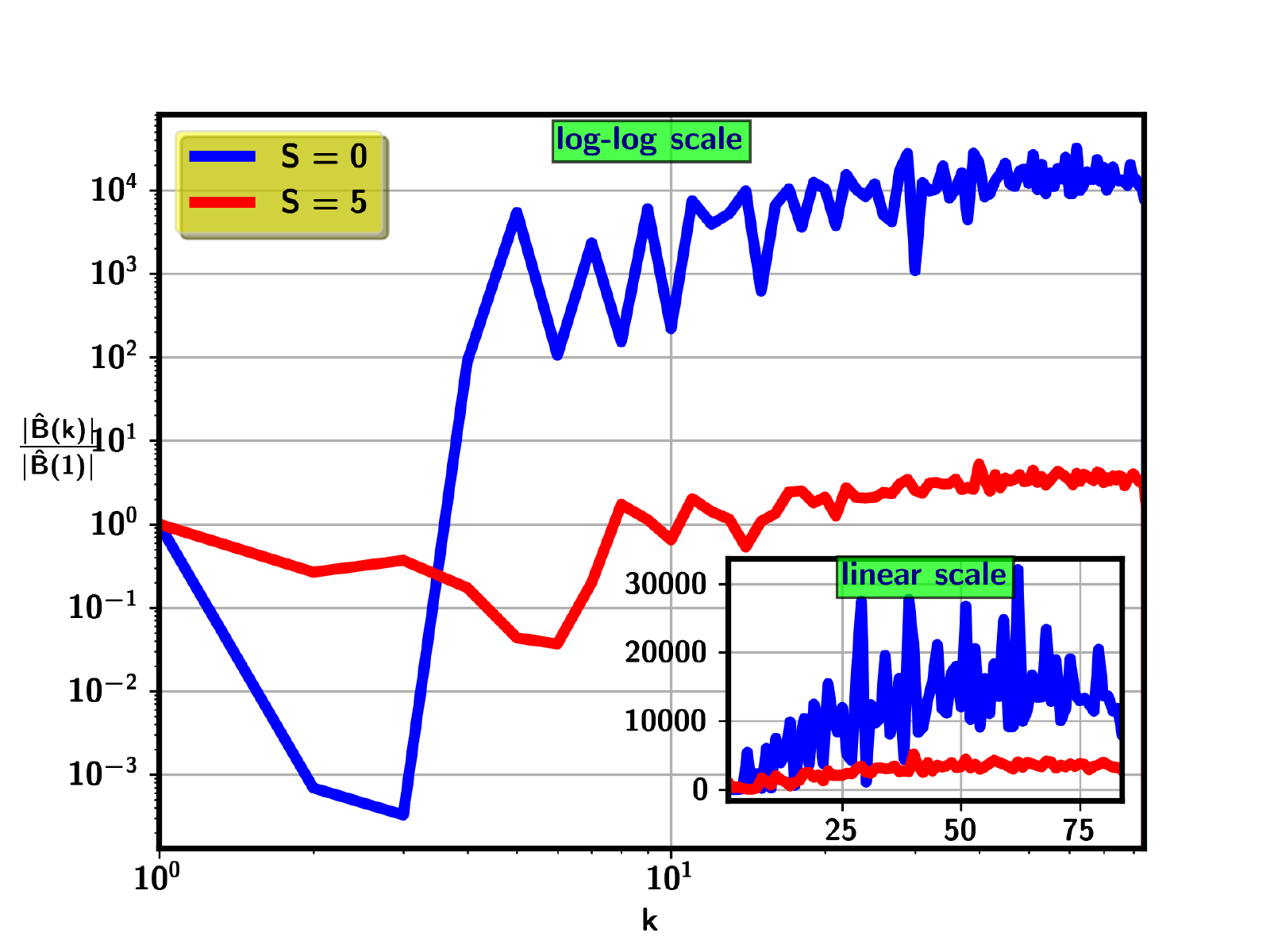}
		\caption{}
	%	\label{ABC spectra}
	\end{subfigure}
	\caption{(a) Magnetic energy ($E_B = \frac{1}{2} \int_{V} (B_x^2 + B_y^2 + B_z^2) dx dy dz$) growth rate ($\gamma = \frac{d}{dt}(\ln E_B(t))$) at late times (eg. $t\sim80$ to $90$) for small scale helical ABC-like flow  in the absence of shear flow ($S = 0$). (b) Magnetic energy iso-surface in the absence of shear flow, $S = 0$ (magnetic energy is effectively dominated by small scales, hence a small scale dynamo (SSD)) [See \textbf{Movie1.mp4}]. (c) Calculation of magnetic energy spectral density (for $S = 0$ \& $5$) $|\hat{B}(k)|$ (such that $\int |\hat{B}(k,t)|^2 dk$ is the total energy at time t (inset view: in linear scale). Simulation details: grid resolution $256^3$.} 
		\label{ABC type}
\end{figure*}

In this study, we use a kinematic dynamo model and begin with a small scale ($k_0 = 8.0$) ABC-like flow (or YM flow with $\beta = 1.0$) to initiate our numerical experiment (See Fig. \ref{initial ABC flow}a).  We perform our numerical runs for a wide range of magnetic Reynolds number $R_m$ and compute the growth rate ($\gamma = \frac{d}{dt}(\ln E_B(t))$) of magnetic energy ($E_B = \frac{1}{2} \int_{V} (B_x^2 + B_y^2 + B_z^2) dx dy dz$) at late times (eg. $t\sim80$ to $90$). For sufficiently large values of the magnetic Reynolds number $R_m$, it is clear from Fig. \ref{ABC type}a that the growth rate of magnetic energy does not saturate with $R_m$, a hallmark of fast dynamo action. It is well-known that magnetic field lines can stretch, twist, and fold (abbreviated as STF) \cite{childress_STF:1995} in a kinematic dynamo model with ABC-like as the base flow. It can be seen that the magnetic energy is concentrated on smaller scales, which can be compared to the length scale of the flow that is driving it (See Fig. \ref{ABC type}b). We have computed the magnetic energy spectral density $|\hat{B}(k)|$ (such that $\int |\hat{B}(k,t)|^2 dk$ is the total energy at time t and $k = \sqrt{k_x^2 + k_y^2 + k_z^2}$). 
From our spectral analysis, we observe that the majority of the power is concentrated in higher modes (i.e., on smaller length scales) (See fig. \ref{ABC type}c). To put it another way, the dynamo is essentially a small scale or fluctuation dynamo (SSD). Investigating the effect of flow shear on this highly helical and chaotic 3D ABC-like flow is an interesting line of inquiry.

Several authors \cite{Brandenburg_MNRAS:2001, Kapyla_APJ:2009, Kim_APJ:2009, Tobias_Nature:2013, Cattaneo_APJ:2014} have addressed the impact of large-scale shear on helical base flows in the context of dynamo action. \textcolor{black}{In some} of the earlier works have considered a circularly polarized, time dependent $2.5$-dimensional base flow namely GP flow \cite{Galloway_Proctor:1992} as a driver for dynamo simulation. In essence, the time dependency of GP flow introduces the chaoticity and stretching property into the system, both of which are necessary for dynamo activity. Another important property of GP flow is that, one can control its reflectional symmetry (helicity distribution) by varying certain physical parameter \cite{Tobias_Nature:2013} similar to our $\beta$ parameter in YM flow \cite{EPI2D:2017, Biswas_arxiv:2022}.

\begin{figure*}
	\begin{subfigure}{0.49\textwidth}
		\centering
		\includegraphics[scale=0.6]{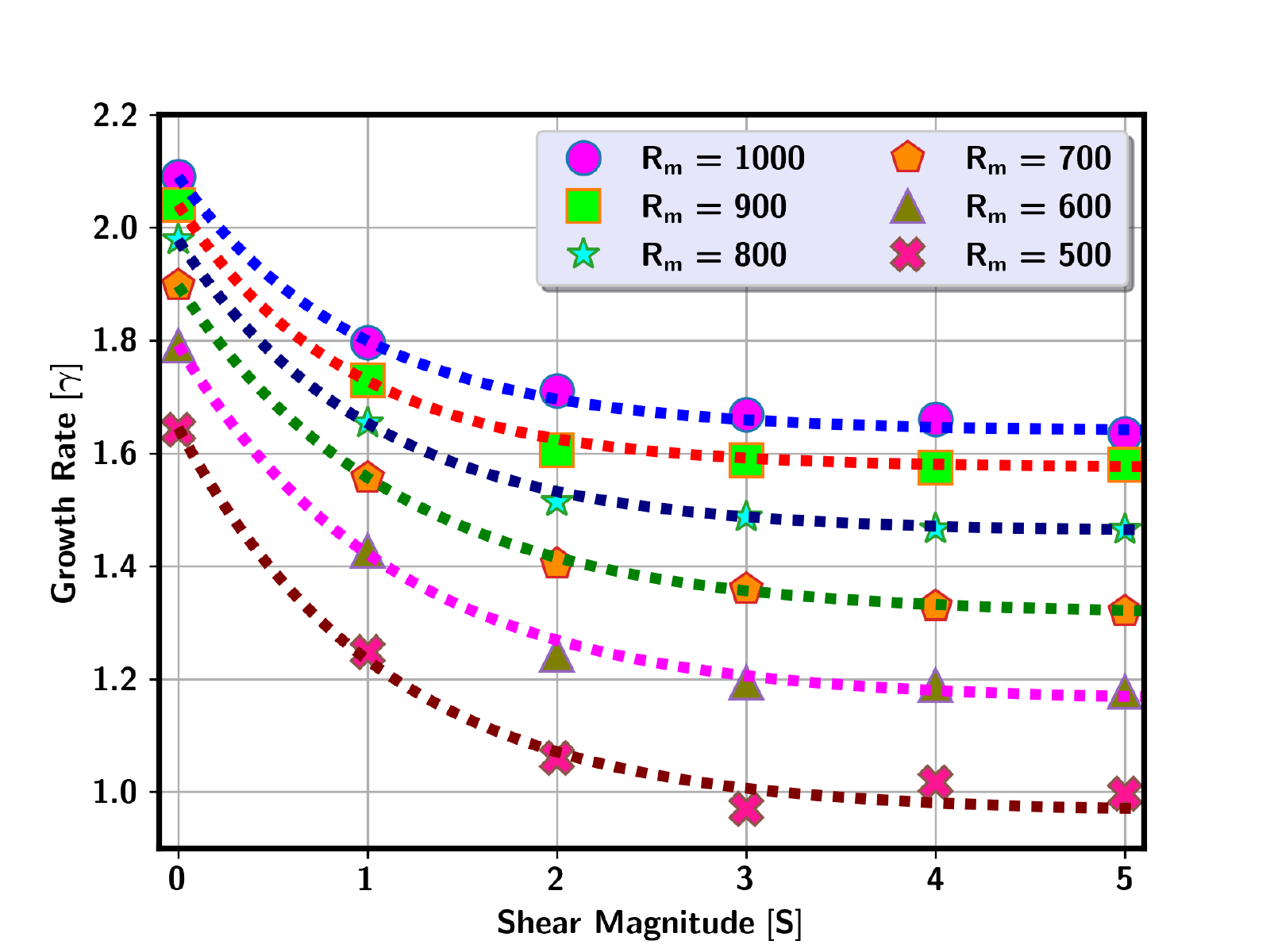}
	%	\caption{}
	\end{subfigure}
 \caption{Magnetic energy ($E_B = \frac{1}{2} \int_{V} (B_x^2 + B_y^2 + B_z^2) dx dy dz$) growth rate ($\gamma = \frac{d}{dt}(\ln E_B(t))$) as function of shear flow strength ($S$) for a helical base flow (ABC-like (i.e, $\beta = 1.0$) flow). As the shear strength ($S$) increases, \textcolor{black}{small} scale magnetic energy growth rate decreases.  The profile of magnetic energy iso-surface (for $S = 0$) and magnetic energy spectral density (for $S = 0$ \& $S = 5$) are shown in Fig. \ref{ABC type}}
 	\label{ABC flow with shear growth rate}
\end{figure*}

Here, using a kinematic dynamo model, we examine the impact of large scale ($k_s = 1$) flow shear (Eq. \ref{cos shear profile}) on small-scale ($k_0 = 8$) 3-dimensional YM flow with $\beta=1$ (i.e, ABC-like flow) (Eq. \ref{ABC_like}) (See Fig. \ref{ABC flow with shear growth rate}). As part of our study, we have conducted numerical simulations with varying shear flow strengths $S$. As can be seen in Fig. \ref{ABC flow with shear growth rate}, the small-scale dynamo action is suppressed by the large-scale ($k_s = 1$) shear flow. The magnetic energy growth rate is found to decrease over a broad range of magnetic Reynolds numbers $R_m$. The interaction between helical base flows and large scale shear effectively limits the growth of small scales (i.e, fluctuations). A possible reason could that in a fully chaotic system, two neighboring fluid element would diverge exponentially in time. If one includes a regulating flow (shear flow), the two neighboring fluid elements will diverge, but algebraically - which implies less chaotic flow and hence reduced dynamo growth \cite{Cattaneo_APJ:2014}. The primary function of flow shear is to diminish the efficacy of fast dynamo action at small scales, which in turn may be interpreted as the flow shear is effectively helping to boost the activity of dynamos at larger scales, or the mean field. This possibility is also reported by a number of authors \cite{Kim_PRE:2009, Tobias_Nature:2013, Cattaneo_APJ:2014, Nigro_MNRAS:2017} in the past, for 2.5 dimensional helical GP flow. Hence our findings in $\beta=1$ limit of YM flow which corroborate with earlier work on helical flows, may be regarded as a benchmark for GMHD3D solver. In light of this background, a reasonable question to ask is the following: what kind of effect does shear flows have on a small-scale base flow that is not helical and does the scale of flow shear matter at all?

\begin{figure*}
	\centering
	\begin{subfigure}{0.49\textwidth}
		\centering
		%\hfill
		\includegraphics[scale=0.6]{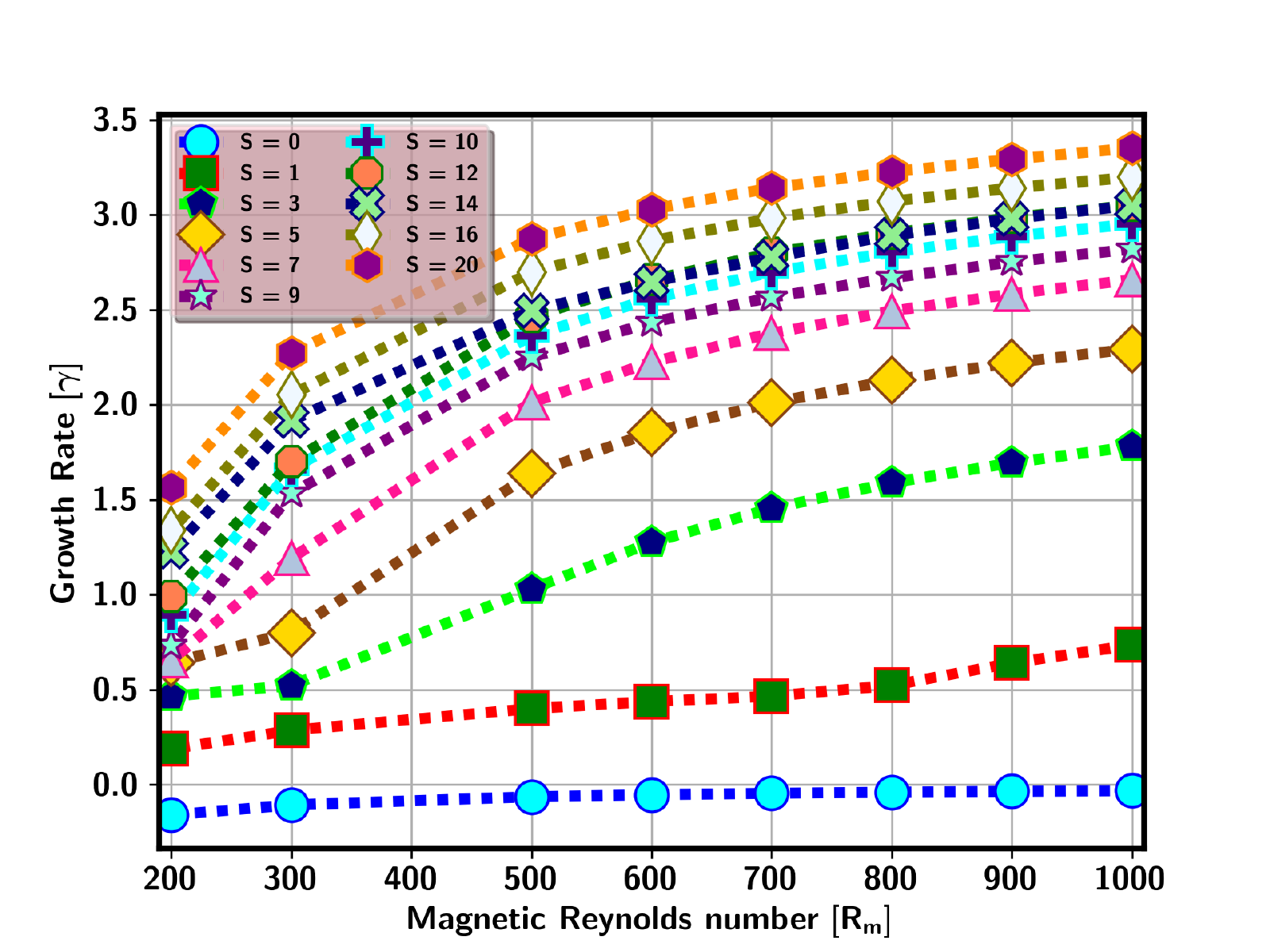}
		\caption{}
	%	\label{EPI2D flow growth rate}
	\end{subfigure}
	\begin{subfigure}{0.49\textwidth}
		\centering
		\includegraphics[scale=0.6]{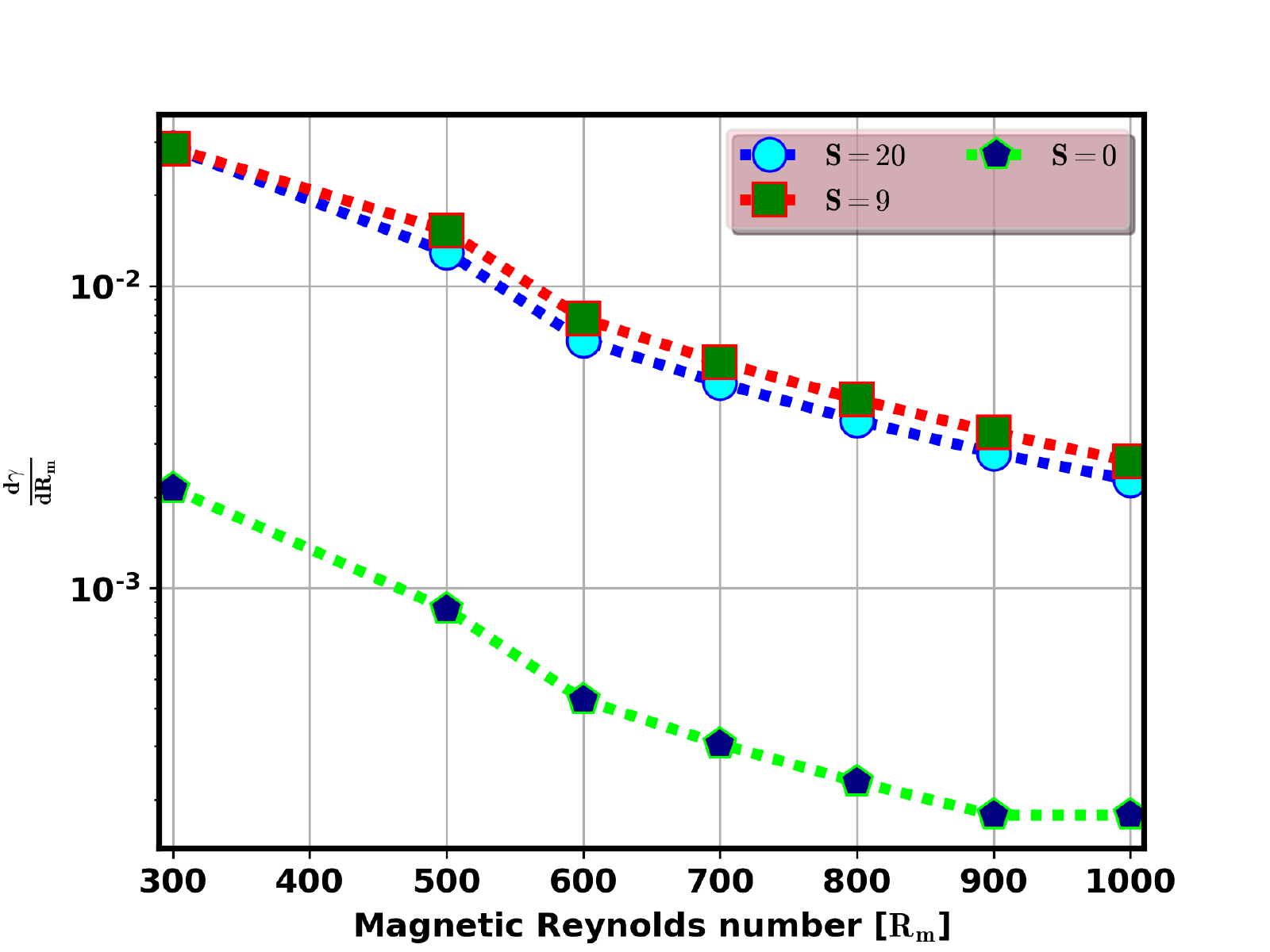}
		\caption{}
%		\label{EPI2D Spectra}
	\end{subfigure}
	\begin{subfigure}{0.49\textwidth}
		\centering
		\includegraphics[scale=0.6]{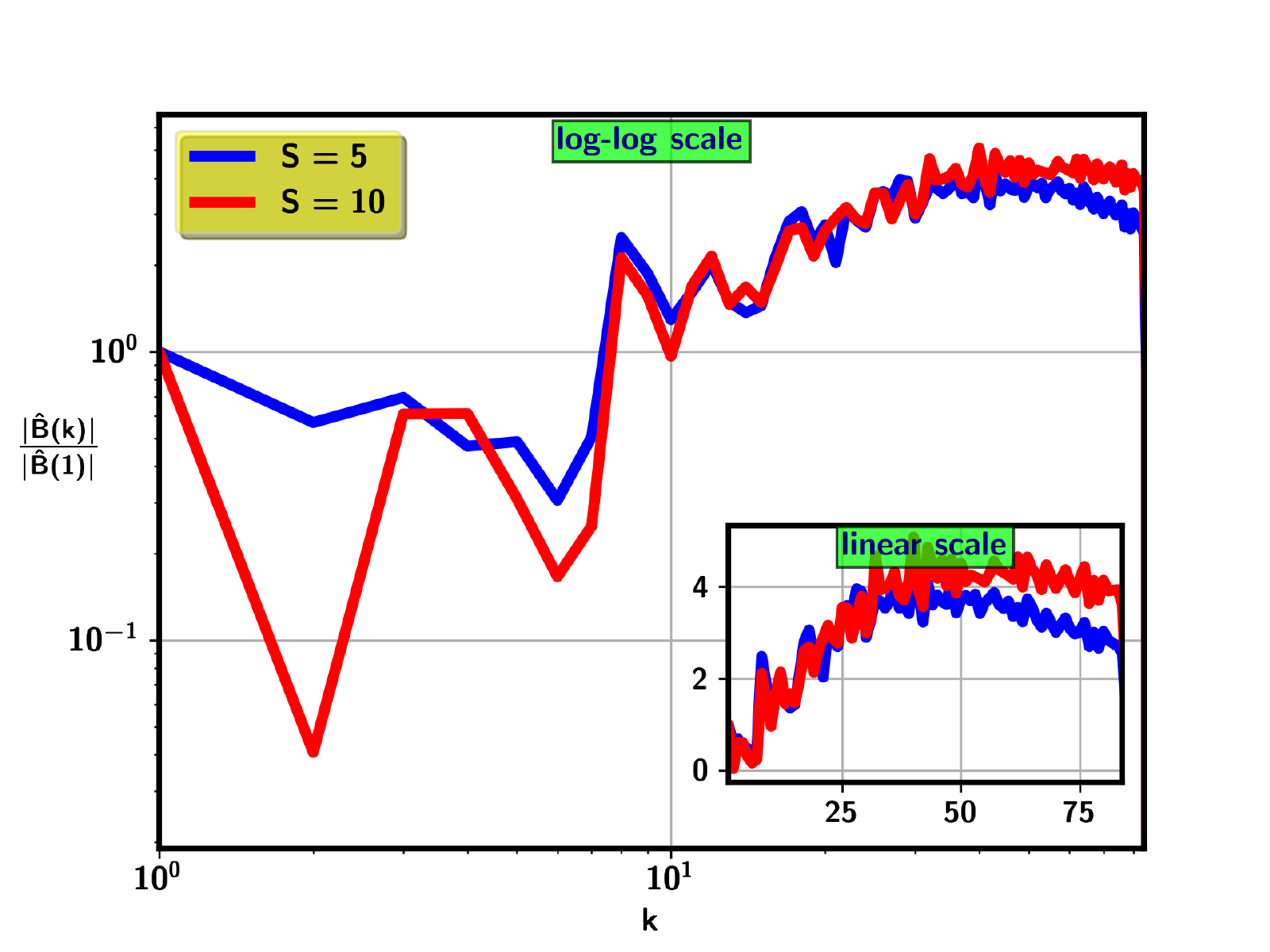}
		\caption{}
%		\label{EPI2D Energy in each mode}
	\end{subfigure}
	\begin{subfigure}{0.49\textwidth}
		\centering
		\includegraphics[scale=0.6]{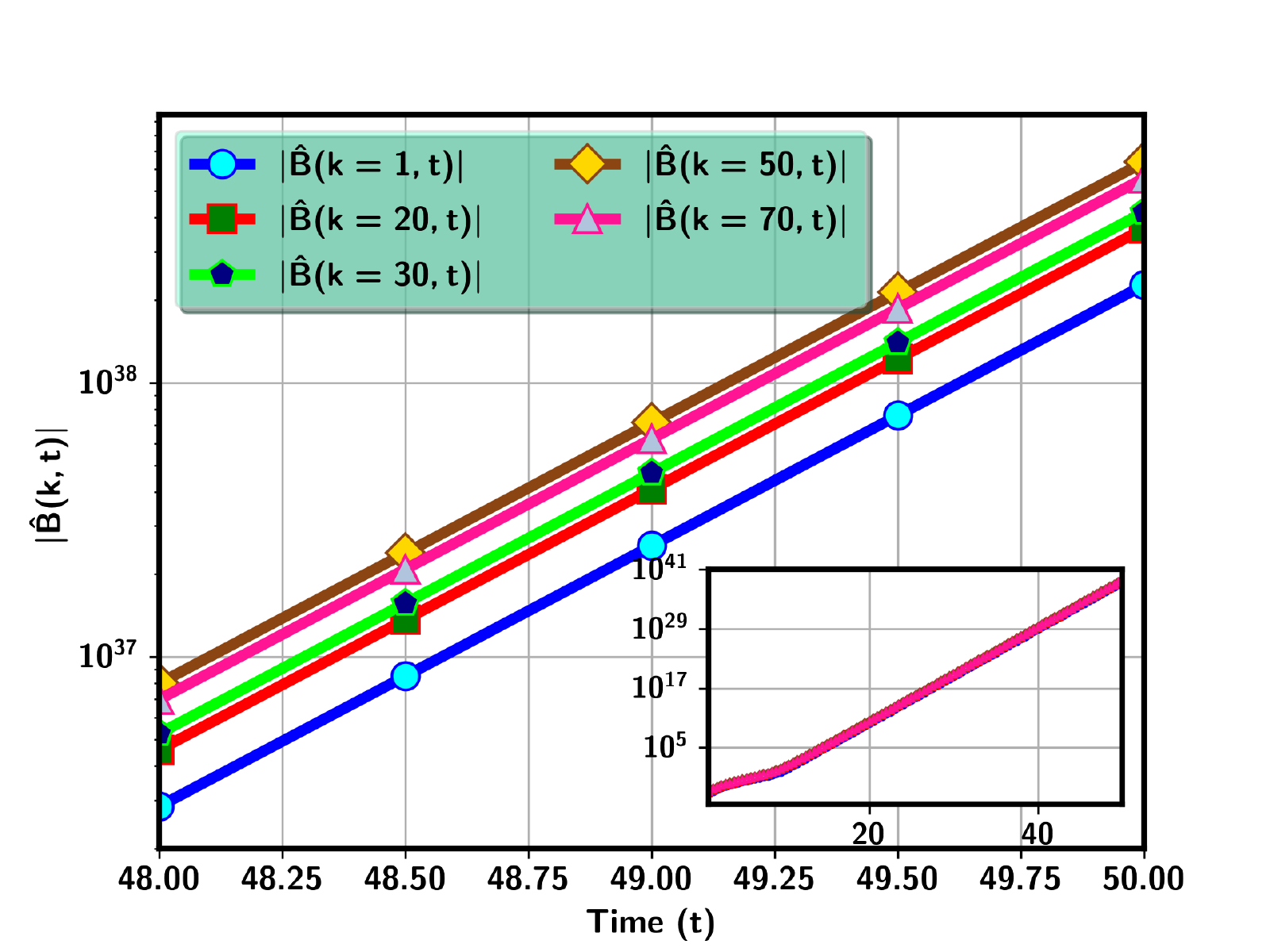}
		\caption{}
	%	\label{EPI2D Energy in each mode}
	\end{subfigure}
	\caption{(a) Magnetic energy ($E_B = \frac{1}{2} \int_{V} (B_x^2 + B_y^2 + B_z^2) dx dy dz$) growth rate ($\gamma = \frac{d}{dt}(\ln E_B(t))$) at late times (eg. $t\sim80$ to $90$) as a function of magnetic Reynolds number $R_m$ for small scale ($k_0 = 8$) non-helical EPI2D flow and for various values of $S$, the large scale ($k_s = 1$) periodic shear flow strength. \textcolor{black}{(b) Calculation of $\frac{d \gamma}{dR_m}$ as a function of magnetic Reynolds number ($R_m$) for different values of shear flow strength $S$.} (c) Calculation of magnetic energy spectral density $|\hat{B}(k)|$ (such that $\int |\hat{B}(k,t)|^2 dk$ is the total energy at time t for two different values of shear flow strength $S$, namely $S = 5$ and $S = 10$ (inset view: in linear scale). (d) Time evolution of magnetic energy spectral density contained in each mode. The higher mode numbers (shorter length scales) contain higher energies at all later times shown; which is a primary characteristic of small scale dynamo (SSD).}
	\label{EPI2D type}
\end{figure*}

%EPI2D cos shear
To address this, we employ direct numerical simulation (DNS) to examine the effect of a superposition of large-scale shear and purely non-helical short-scale EPI2D flow (See Fig. \ref{initial ABC flow}c). When there is no flow shear present, the dynamo effect is absent for an EPI2D flow. The magnetic energy growth rate ($\gamma = \frac{d}{dt}(\ln E_B(t))$) for small scale EPI2D flow in the absence of flow shear, is negative over a wide range of magnetic Reynolds numbers $R m$ (See Fig. \ref{EPI2D type}a for $S=0$). This is an obvious indication of non-dynamo activity. \textcolor{black}{However, Zeldovich's classic anti-dynamo theorem provides an alternative explanation for this \cite{Zeldovich_Anti:1957, Rincon:2019}.} When this small-scale, non-dynamo-producing, EPI2D flow is superposed with a large-scale flow shear, the dynamics is found to undergo an interesting transformation. We have carried out our numerical experiments across a broad range of shear flow strengths ($S$) and magnetic Reynolds numbers ($R_m$) (See Fig. \ref{EPI2D type}a for $S = 1$ to $20$). In the presence of a non-zero shear flow strength, as shown in Fig. \ref{EPI2D type}a, magnetic energy growth is clearly visible. At sufficiently high magnetic Reynolds numbers $R_m$, the growth rate of magnetic energy ($\gamma$) continues to vary significantly and is found not to saturate with $R_m$, verifying one of the defining feature of fast dynamo action \cite{childress_STF:1995, Bouya:2013}. \textcolor{black}{We have calculated $\frac{d \gamma}{dR_m}$ as a function of $R_m$ and find that the value of $\frac{d \gamma}{dR_m}$ to be slowly varying, even at the maximum magnetic Reynolds number [See Fig. \ref{EPI2D type}b].} 

\begin{figure*}
	\centering
	\begin{subfigure}{0.49\textwidth}
		\centering
		%\hfill
		\includegraphics[scale=0.08480]{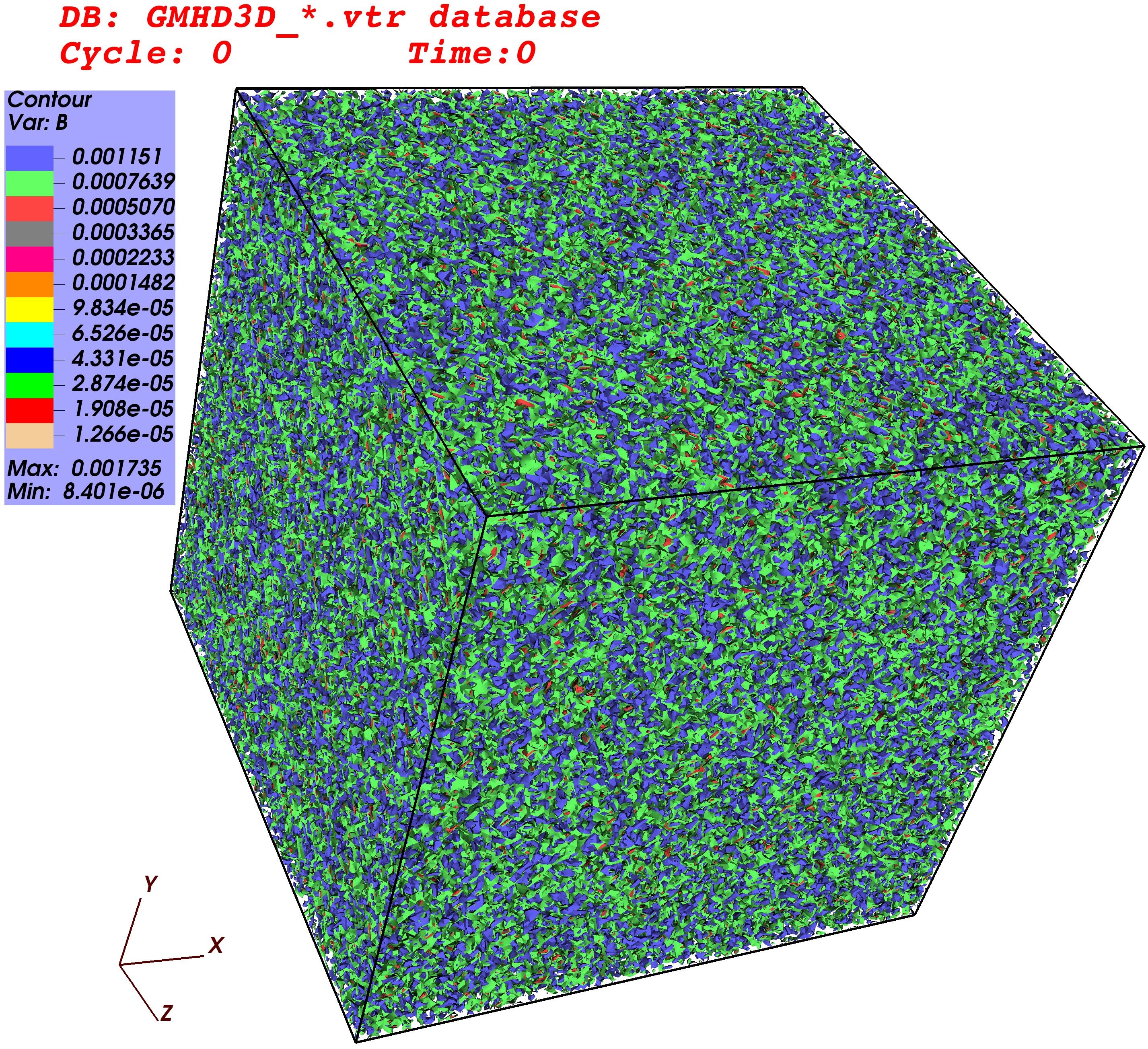}
		\caption{Time = 0.0}
		%	\label{initial flow beta 0}
	\end{subfigure}
	\begin{subfigure}{0.49\textwidth}
		\centering
		\includegraphics[scale=0.08480]{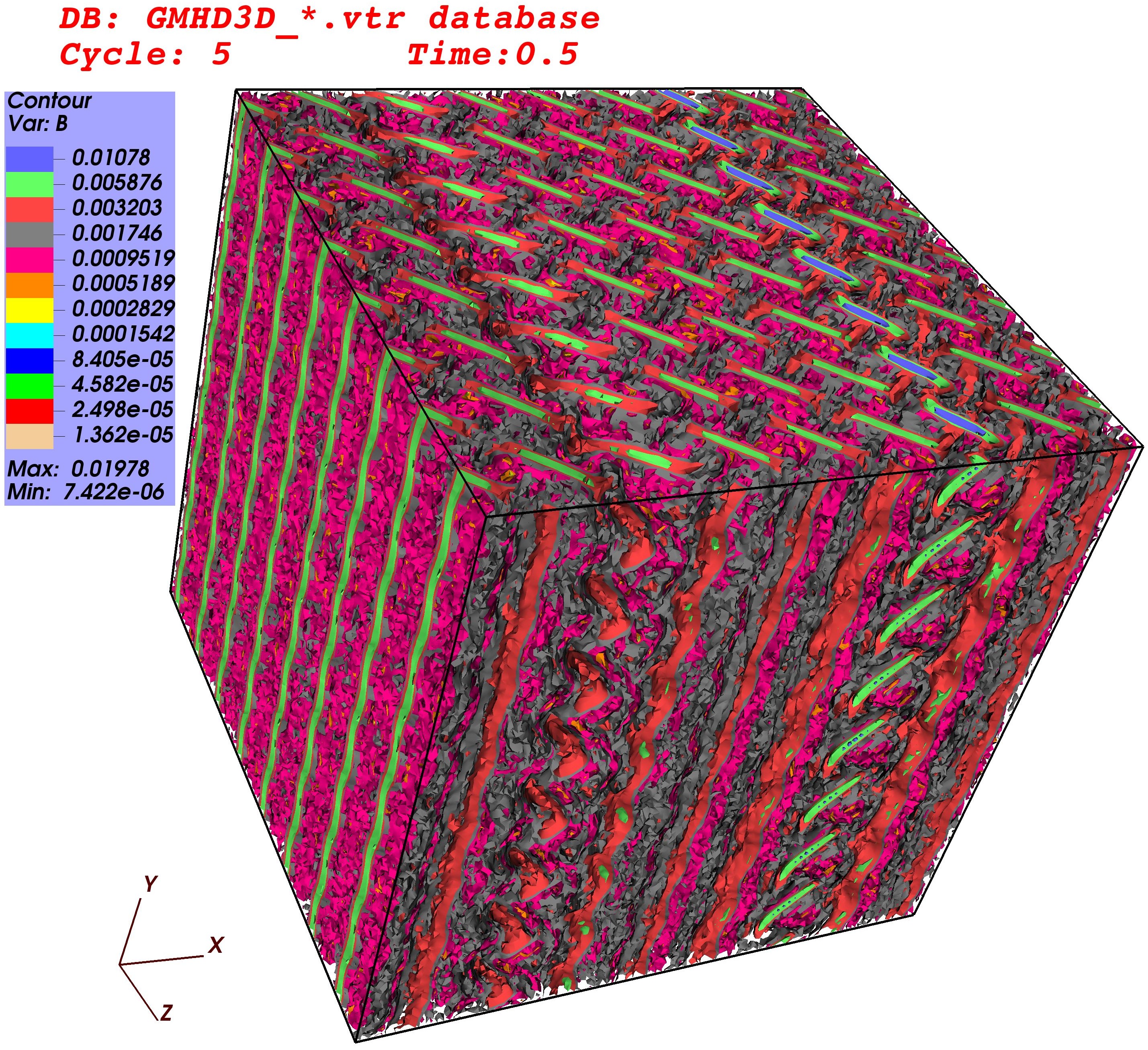}
		\caption{Time = 0.5}
		%\label{initial flow beta 0p4}
	\end{subfigure}
	\begin{subfigure}{0.49\textwidth}
		\centering
		%\hfill
		\includegraphics[scale=0.08480]{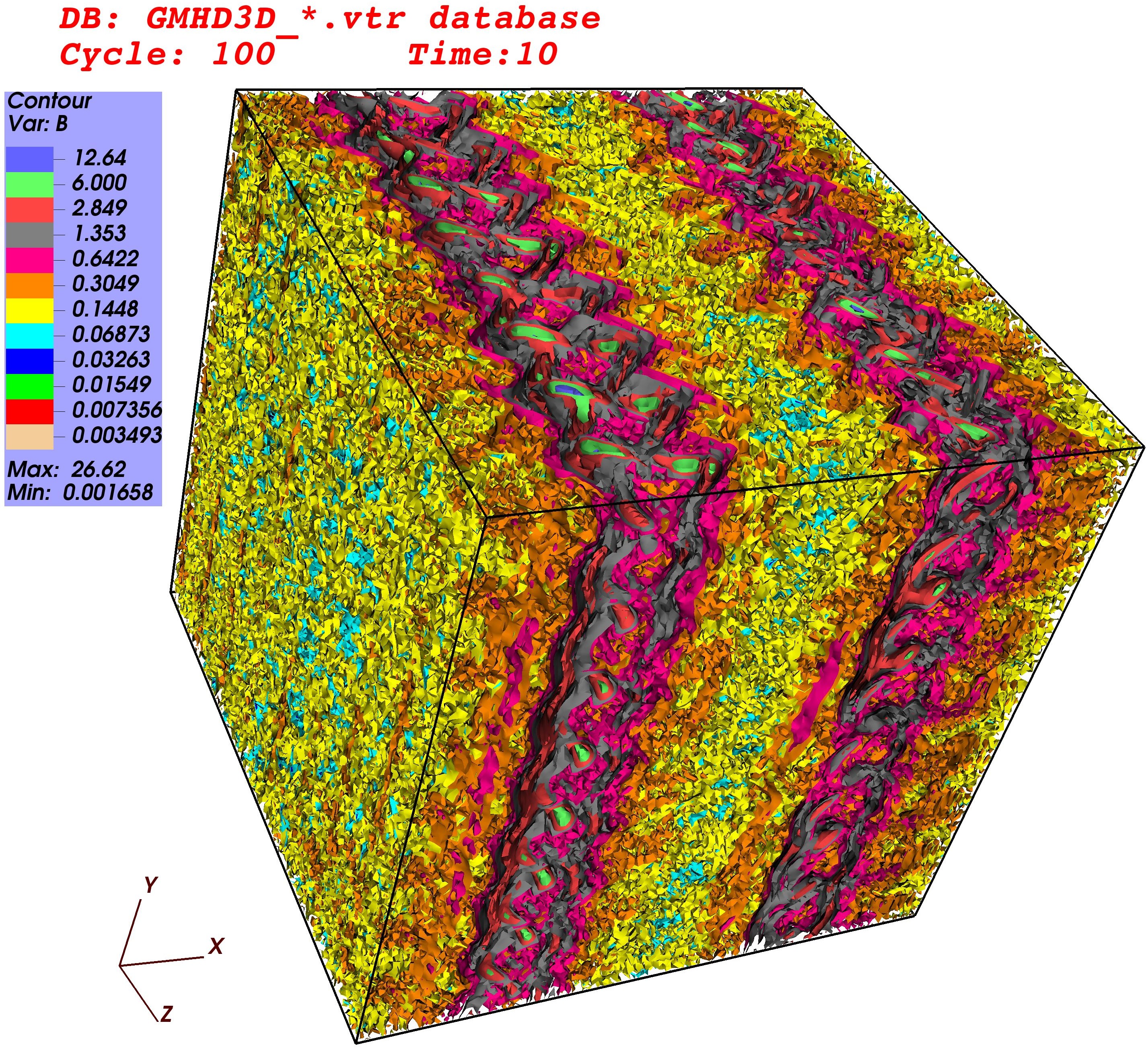}
		\caption{Time = 10.0}
		%	\label{initial flow beta 0}
	\end{subfigure}
	\begin{subfigure}{0.49\textwidth}
		\centering
		\includegraphics[scale=0.08480]{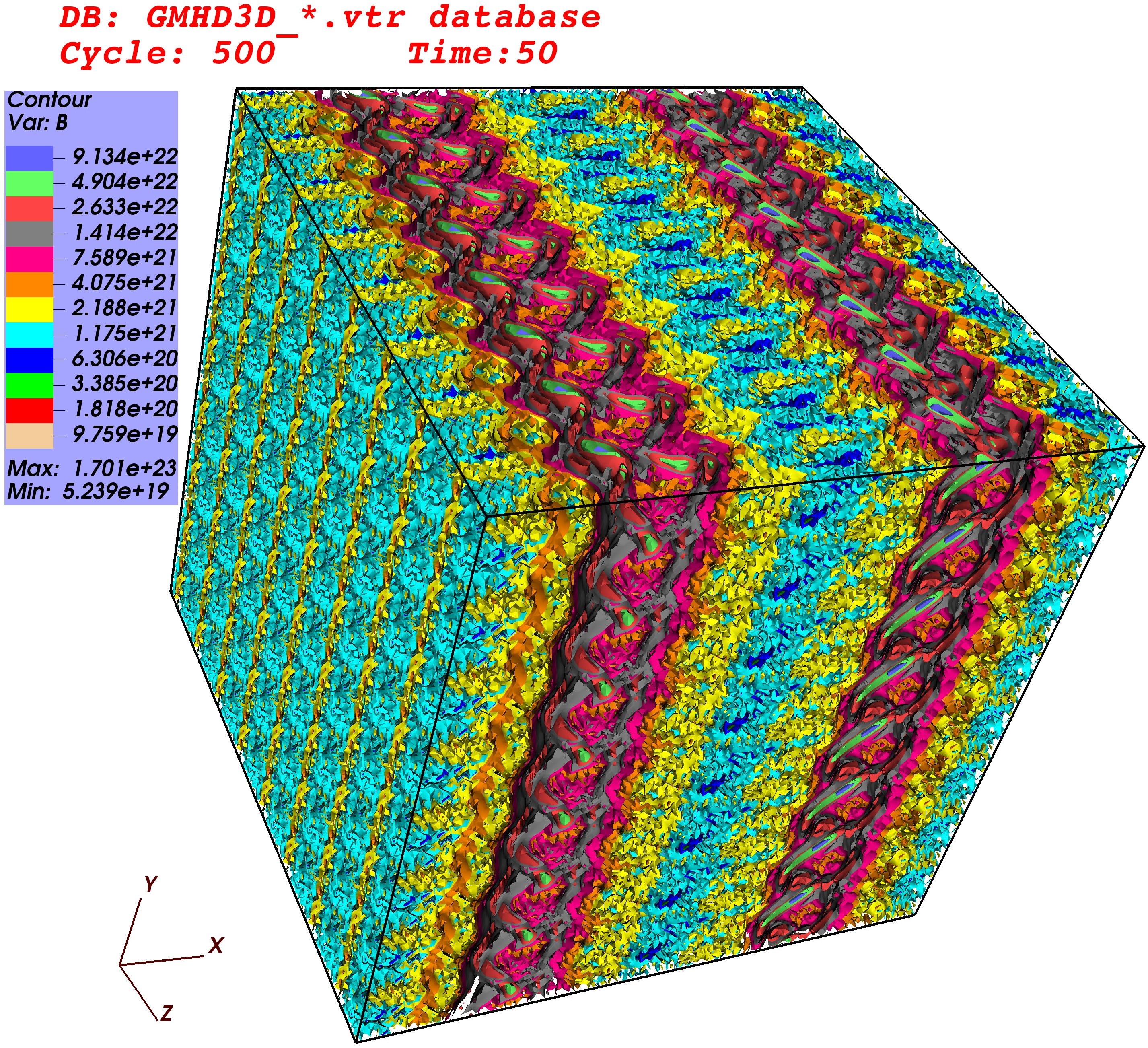}
		\caption{Time = 50.0}
		%\label{initial flow beta 0p4}
	\end{subfigure}
	\caption{Time evolution of 3-dimensional magnetic field iso-surfaces (Iso-B surfaces) for a given small scale ($k_0 = 8$) EPI2D flow and large scale ($k_s = 1$) periodic shear. Dominant magnetic energies are mostly confined in two regimes restricted near the segments ($z = \frac{\pi}{2}$ and $z = \frac{3\pi}{2}$) where the velocity gradients are strongest. The structures are mostly dominated by \textcolor{black}{small} scale structures (SSD) [\textbf{Movie2.mp4}]. Simulation details: grid resolution $256^3$, shear flow strength $S = 5.0$, magnetic Reynolds number $R_m = 200.0$. Visualization in log scale.}
	\label{EPI2D flow B field evolution}
\end{figure*}

We have visualized the iso-surfaces of the magnetic fields (Iso-B surfaces) in three dimensions  and find that the magnetic energy is concentrated in two bands near the segments with strong velocity gradients (See Fig. \ref{EPI2D flow B field evolution}). Magnetic energy iso-surfaces are dominated by \textcolor{black}{small}-scale structures (compared to the length scale of the flow), as is evident from Fig. \ref{EPI2D flow B field evolution}. We have computed the magnetic energy spectral density $|\hat{B}(k)|$ (such that $\int |\hat{B}(k,t)|^2 dk$ is the total energy at time t and $k = \sqrt{k_x^2 + k_y^2 + k_z^2}$) to verify our findings. It is seen from Fig. \ref{EPI2D type}c that the most of the magnetic energy is at the higher mode numbers (shorter length scales). To further illustrate our point, we have plotted the magnetic energy spectral density over time for each of the modes (eg. $|\hat{B}(k = 1), t|, |\hat{B}(k = 20), t|, |\hat{B}(k = 30), t|, |\hat{B}(k = 50), t|, |\hat{B}(k = 70), t|$ etc.). It is easy to see from Fig. \ref{EPI2D type}d that the higher modes (shorter length scales) contain higher energies; this is a primary characteristic of SSD.

\begin{figure*}
	\begin{subfigure}{0.49\textwidth}
		\centering
		\includegraphics[scale=0.6]{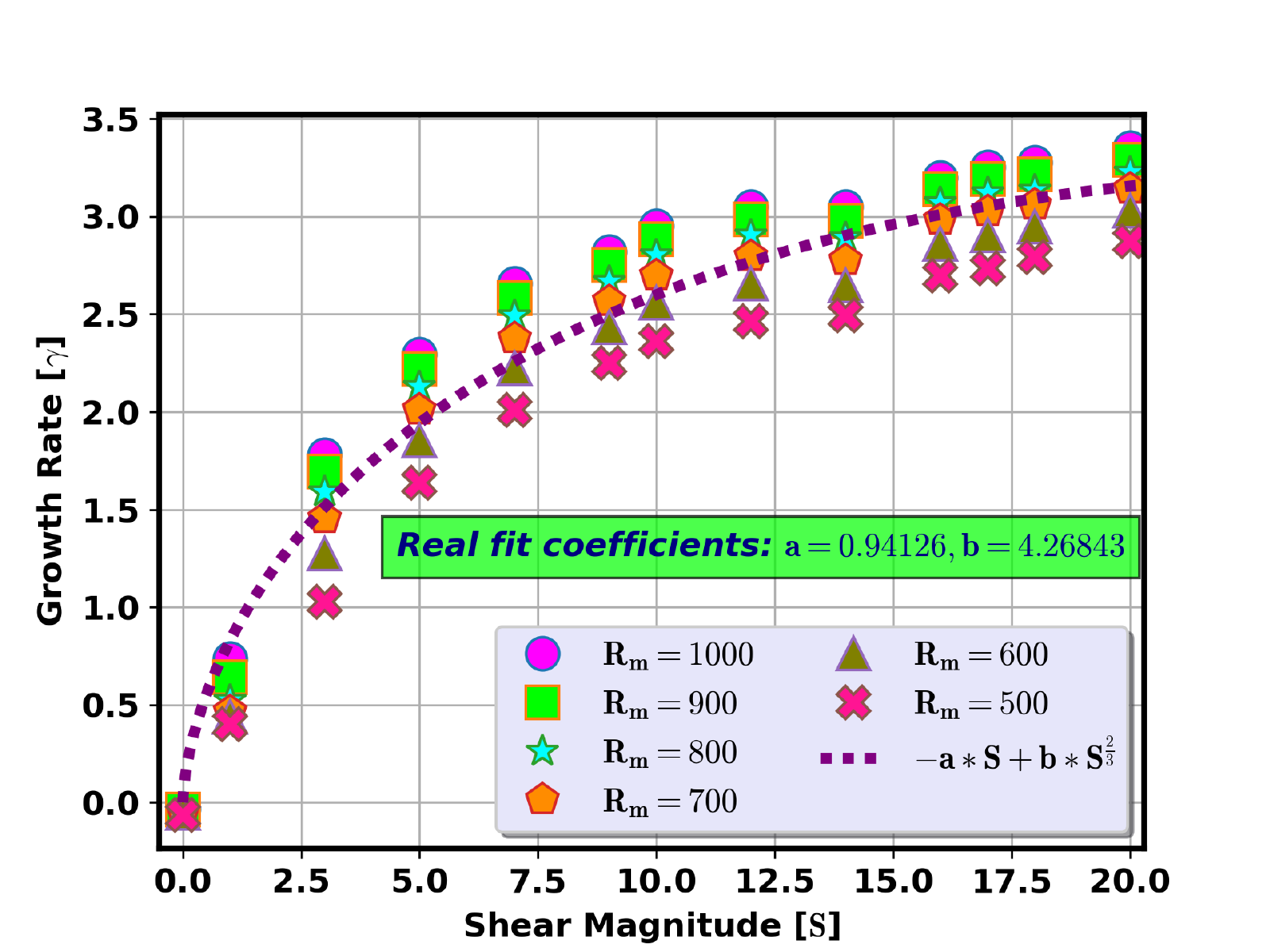}
		%	\caption{}
	\end{subfigure}
	\caption{ Magnetic energy ($E_B = \frac{1}{2} \int_{V} (B_x^2 + B_y^2 + B_z^2) dx dy dz$) growth rate ($\gamma = \frac{d}{dt}(\ln E_B(t))$) as function of shear flow strength ($S$) for small scale ($k_0 = 8$) non-helical EPI2D flow and large scale ($k_s = 1$) periodic shear flow. As the shear strength ($S$) increases, \textcolor{black}{small} scale magnetic energy growth rate also increases following an algebraic scaling of the form $- aS + bS^\frac{2}{3}$ with $a = 0.94126$, $b = 4.26843$.}
%\textcolor{red}{(b) Magnetic energy growth rate ($\gamma$) as function of initial injected fluid helicity ($\beta$) for small scale ($k_0 = 8$) YM flow and large scale ($k_s = 1$) periodic shear flow. As the injected fluid helicity ($\beta$) increases, short scale magnetic energy growth rate is seen to suppressed.}	
\label{EPI2D flow with shear growth rate}	
\end{figure*}

In addition, we plot the rate of increase in magnetic energy ($\gamma$) as a function of the shear flow strength ($S$) for non-helical base flow. Figure \ref{EPI2D flow with shear growth rate} demonstrates unambiguously that as the amplitude of the shear flow increases, the rate of growth of magnetic energy also increases. Spectral analysis confirms the SSD-like structure for a given value of shear flow strength ($S$) (See Fig. \ref{EPI2D type}c). Based on the evidence presented in Fig. \ref{EPI2D flow with shear growth rate}, we conclude that the effect of large-scale shear flows in this instance is not to suppress the SSD activity, but rather to amplify it. Additionally, we obtain a generalized algebraic (combination of linear and non-linear) scaling \textcolor{black}{(based on $\chi^2$ minimization)} for the rate of increase of magnetic energy ($\gamma$) in the form $\gamma (S) = - aS + bS^\frac{2}{3}$ (See Fig. \ref{EPI2D flow with shear growth rate}), where $a$ \& $b$ are real fit coefficients.  Our numerical finding of dependency of $\gamma$ on $S$ is found to be in close agreement with analytic predictions \cite{Kolokolov:2011, Proctor_JFM:2012}. It is observed that, for a given random smooth velocity field, large-scale shear can support an small scale dynamo (SSD) with a scaling of $S^\frac{2}{3}$ \cite{Kolokolov:2011}, which is consistent with an upper bound for growth rates anticipated afterward \cite{Proctor_JFM:2012}.

It has been recently proposed \cite{Singh_APJL:2017} that, a random non-helically driven, dissipative model can enhance SSD action. This work reports on amplification of SSD using a kinematic dynamo model which solves a Navier-Stokes equation for the fluid flow, along with a linear background shear and a random non-helical white noise drive \cite{Singh_APJL:2017}. In yet another model, the shear is self-consistently driven by the presence of an in-plane temperature gradient resulting in SSD \cite{Currie:2019}.

In contrast to these earlier studies, in our model we have imposed a driven shear velocity field with a three-dimensional flow field (the helicity of which can be controlled) and studied the growth of magnetic energy by only evolving the induction equation for magnetic fields in a triply periodic three-dimensional box (i.e, the velocity field is not evolved using the Navier-Stokes equation; rather, it is given and remains static throughout the simulation.). Using this configuration, we have demonstrated unambiguously how flow shears enhance small scale dynamo (SSD) activity for non-helical base flows.

\begin{figure*}
	\centering
	\begin{subfigure}{0.49\textwidth}
		\centering
		%\hfill
		\includegraphics[scale = 0.6]{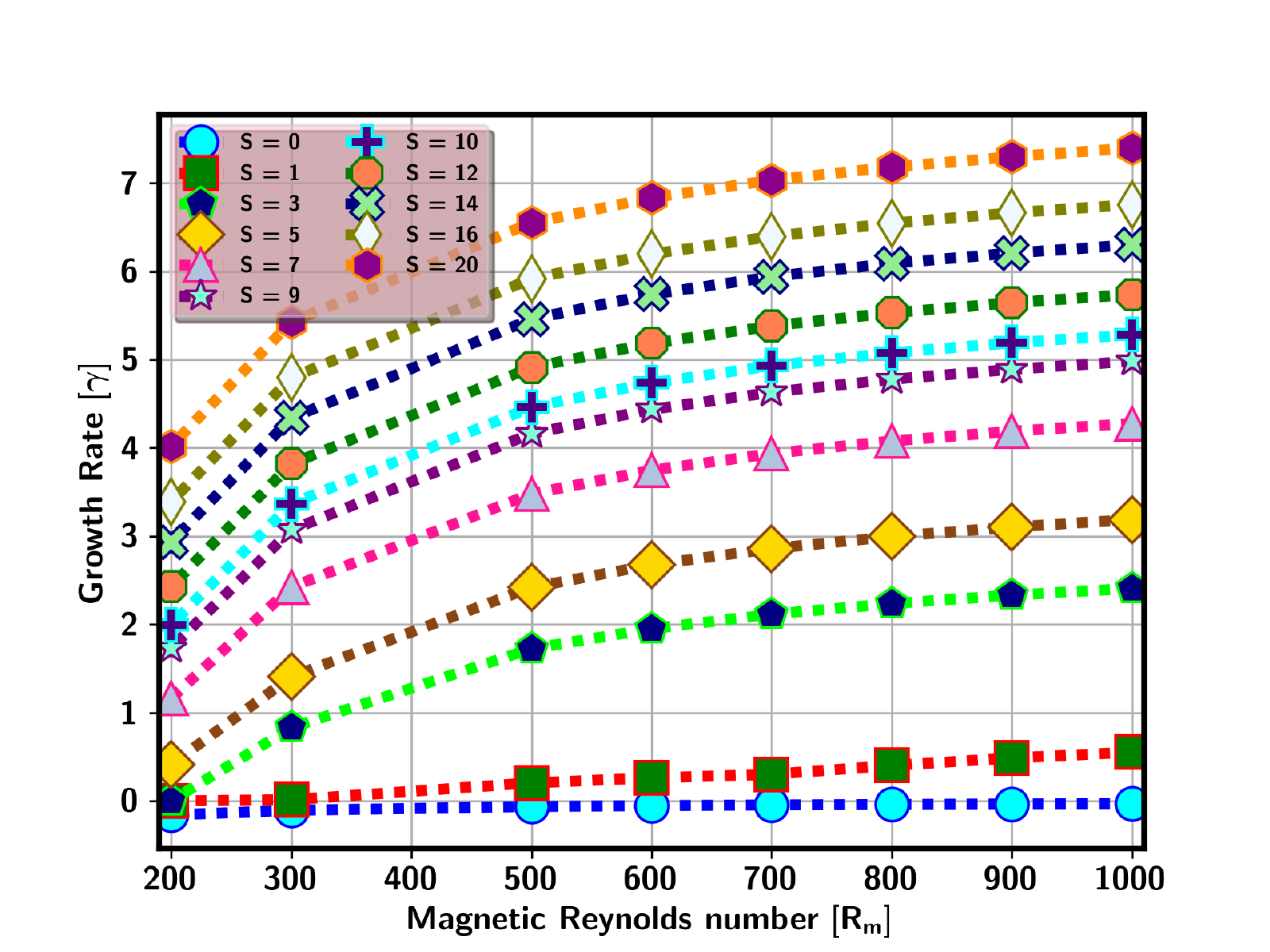}
		\caption{}
	%	\label{Strip Shear Growth Rate}
	\end{subfigure}
	\begin{subfigure}{0.49\textwidth}
		\centering
		\includegraphics[scale = 0.6]{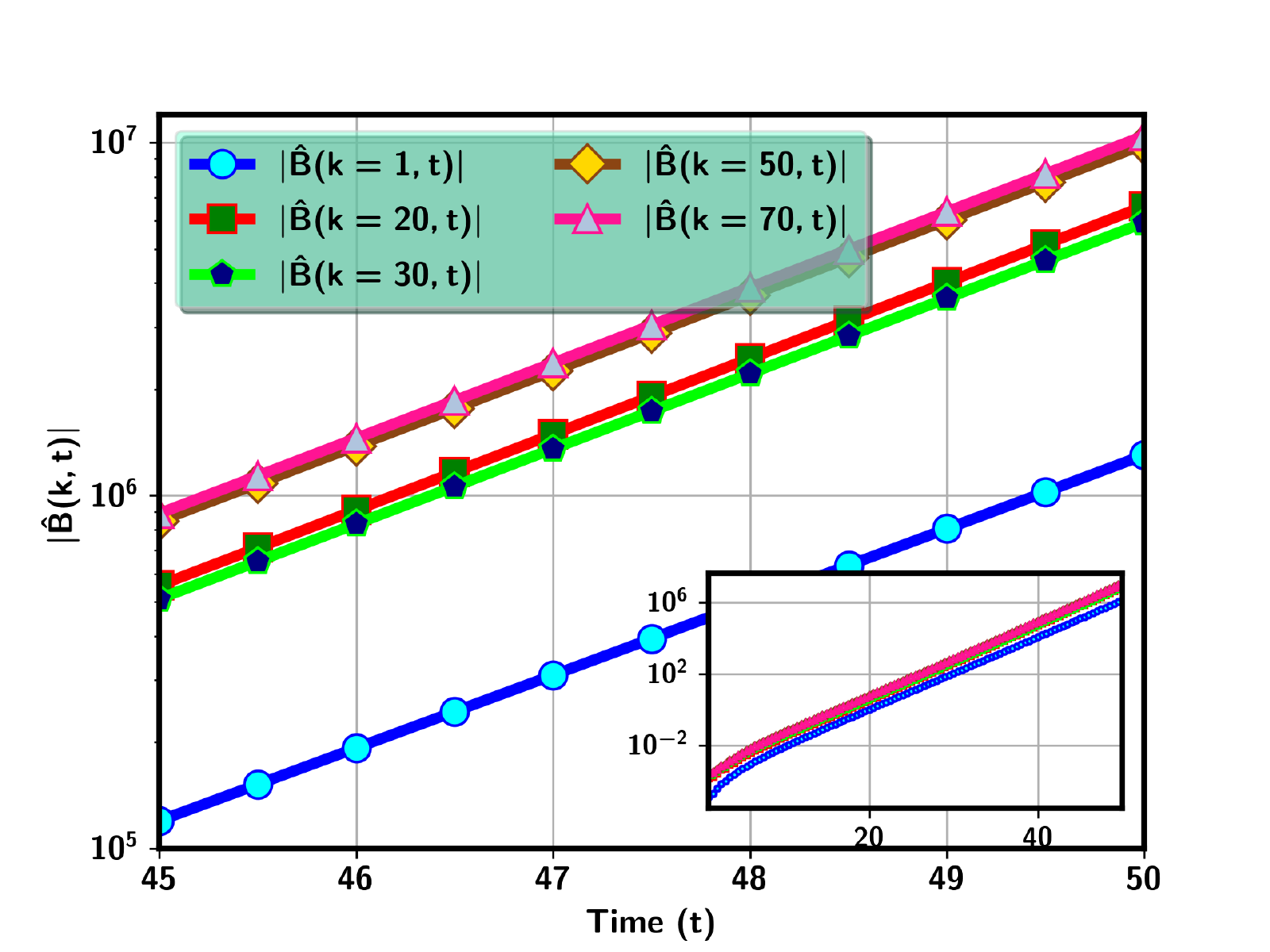}
		\caption{}
	%	\label{Strip Spectra}
	\end{subfigure}
	\begin{subfigure}{0.49\textwidth}
		\centering
		\includegraphics[scale = 0.6]{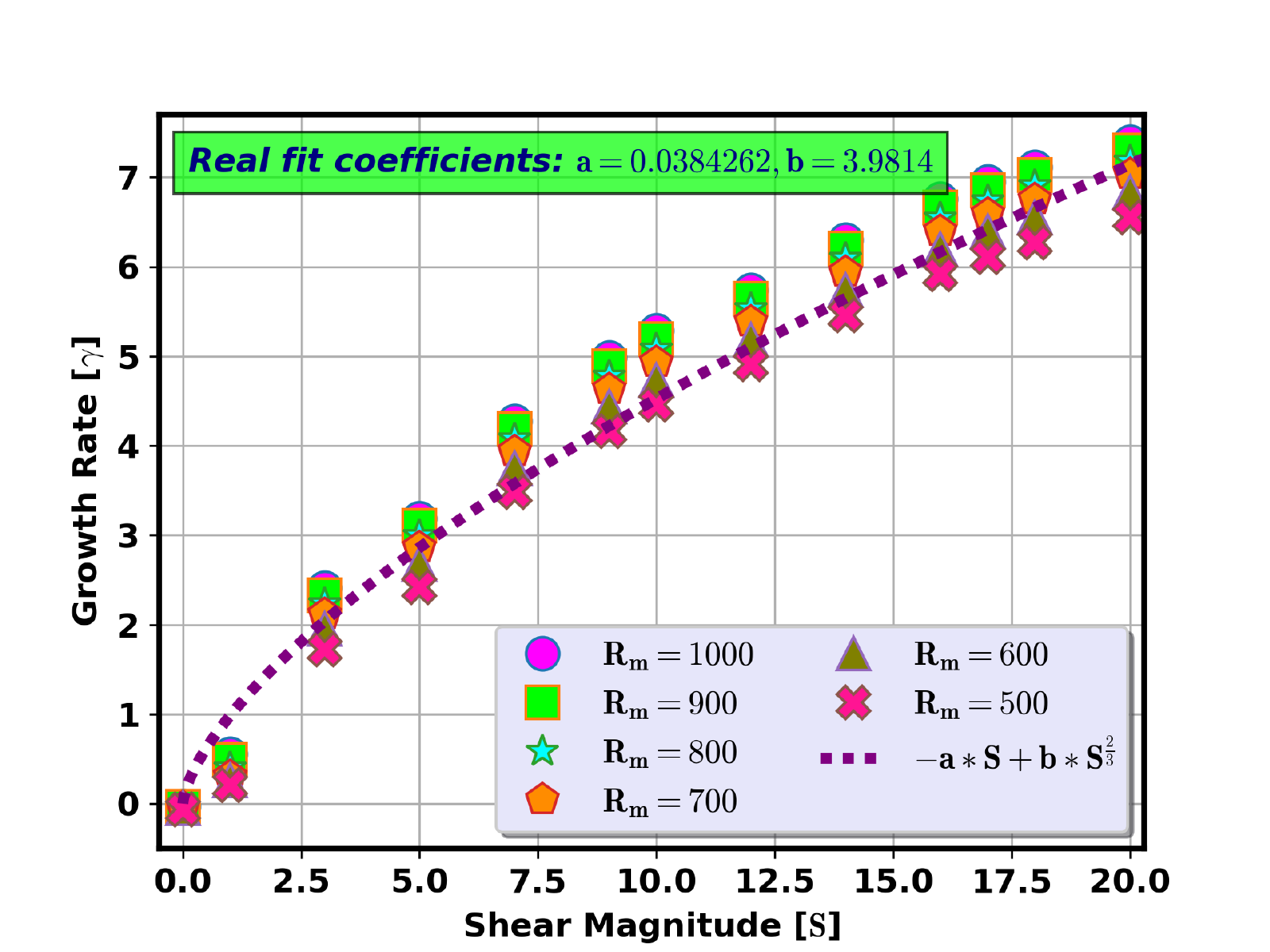}
		\caption{}
	%	\label{Strip Scaling}
	\end{subfigure}
	\caption{(a) Magnetic energy ($E_B = \frac{1}{2} \int_{V} (B_x^2 + B_y^2 + B_z^2) dx dy dz$) growth rate ($\gamma = \frac{d}{dt}(\ln E_B(t))$) at late times (eg. $t\sim80$ to $90$) as a function of magnetic Reynolds number $R_m$ for small scale ($k_0 = 8$) non-helical EPI2D flow and broken jet flow shear ($k_s \to \infty$. i.e, $k_s \sim k_{max}$ numerically speaking). (b) Time evolution of magnetic energy spectral density ($|\hat{B}(k, t)|$) contained in each mode. (c) Magnetic energy ($E_B = \frac{1}{2} \int_{V} (B_x^2 + B_y^2 + B_z^2) dx dy dz$) growth rate ($\gamma = \frac{d}{dt}(\ln E_B(t))$) as function of shear flow strength ($S$) for small scale ($k_0 = 8$) non-helical EPI2D flow and broken jet flow shear ($k_s \to \infty$). As the shear strength ($S$) increases, \textcolor{black}{small} scale magnetic energy growth rate also increases following an algebraic scaling of the form $- aS + bS^\frac{2}{3}$, with $a = 0.0384262$, $b = 3.9814$.}
	\label{EPI2D Strip type}
\end{figure*}

\begin{figure*}
	\centering
	\begin{subfigure}{0.49\textwidth}
		\centering
		%\hfill
		\includegraphics[scale=0.08480]{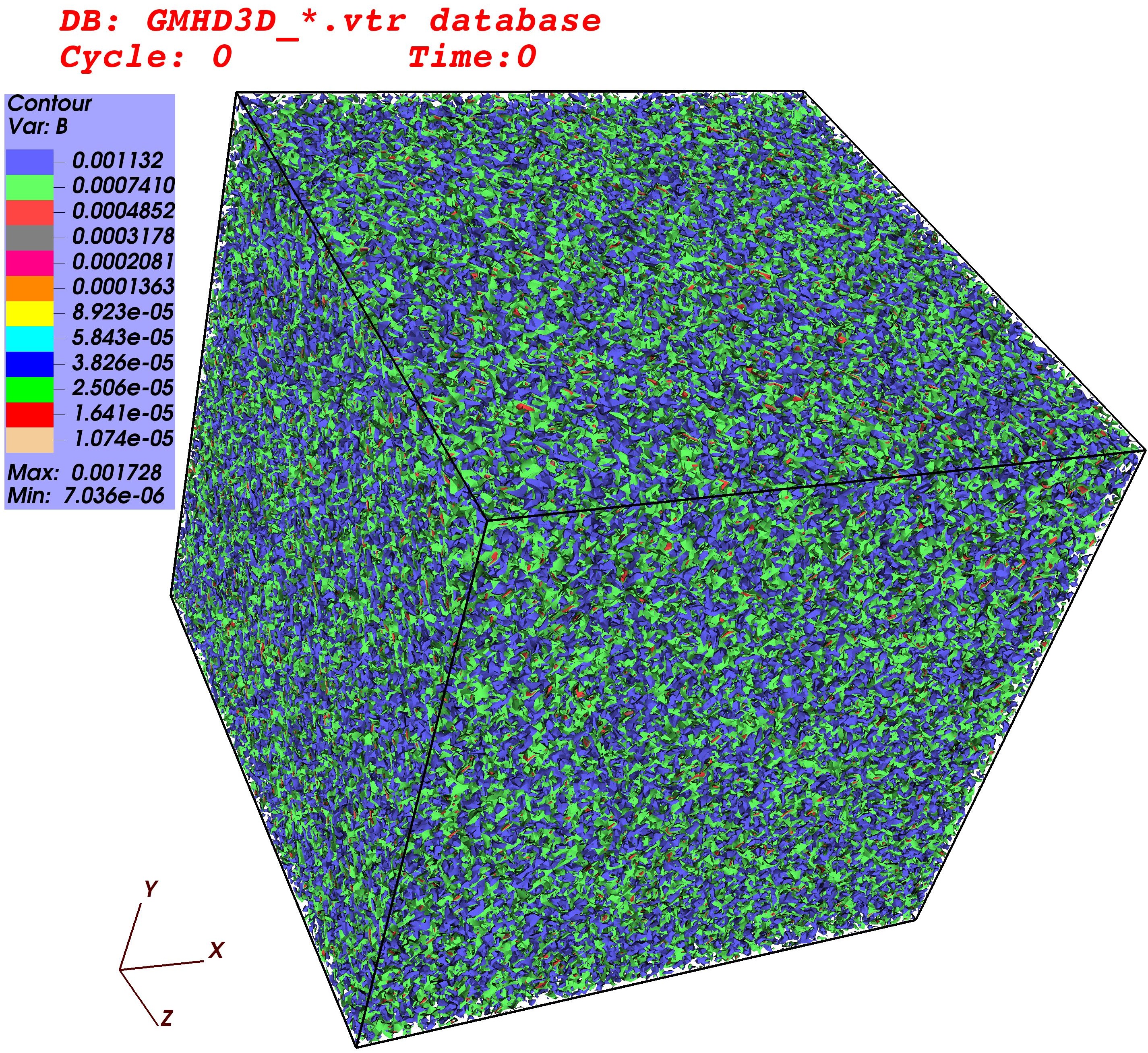}
		\caption{Time = 0.0}
		%	\label{initial flow beta 0}
	\end{subfigure}
	\begin{subfigure}{0.49\textwidth}
		\centering
		\includegraphics[scale=0.08480]{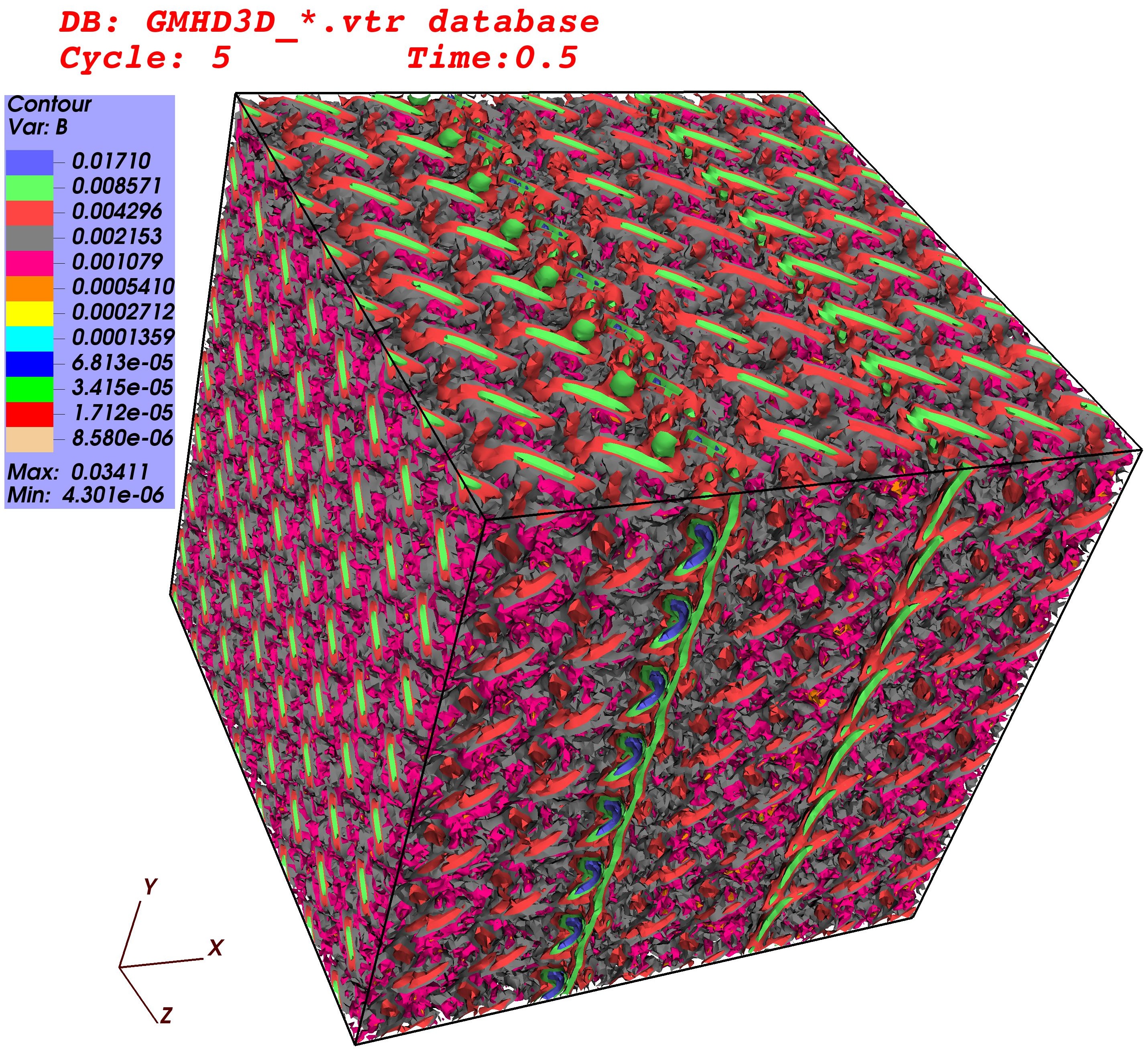}
		\caption{Time = 0.5}
		%	\label{initial flow beta 0p2}
	\end{subfigure}
	\begin{subfigure}{0.49\textwidth}
		\centering
		\includegraphics[scale=0.08480]{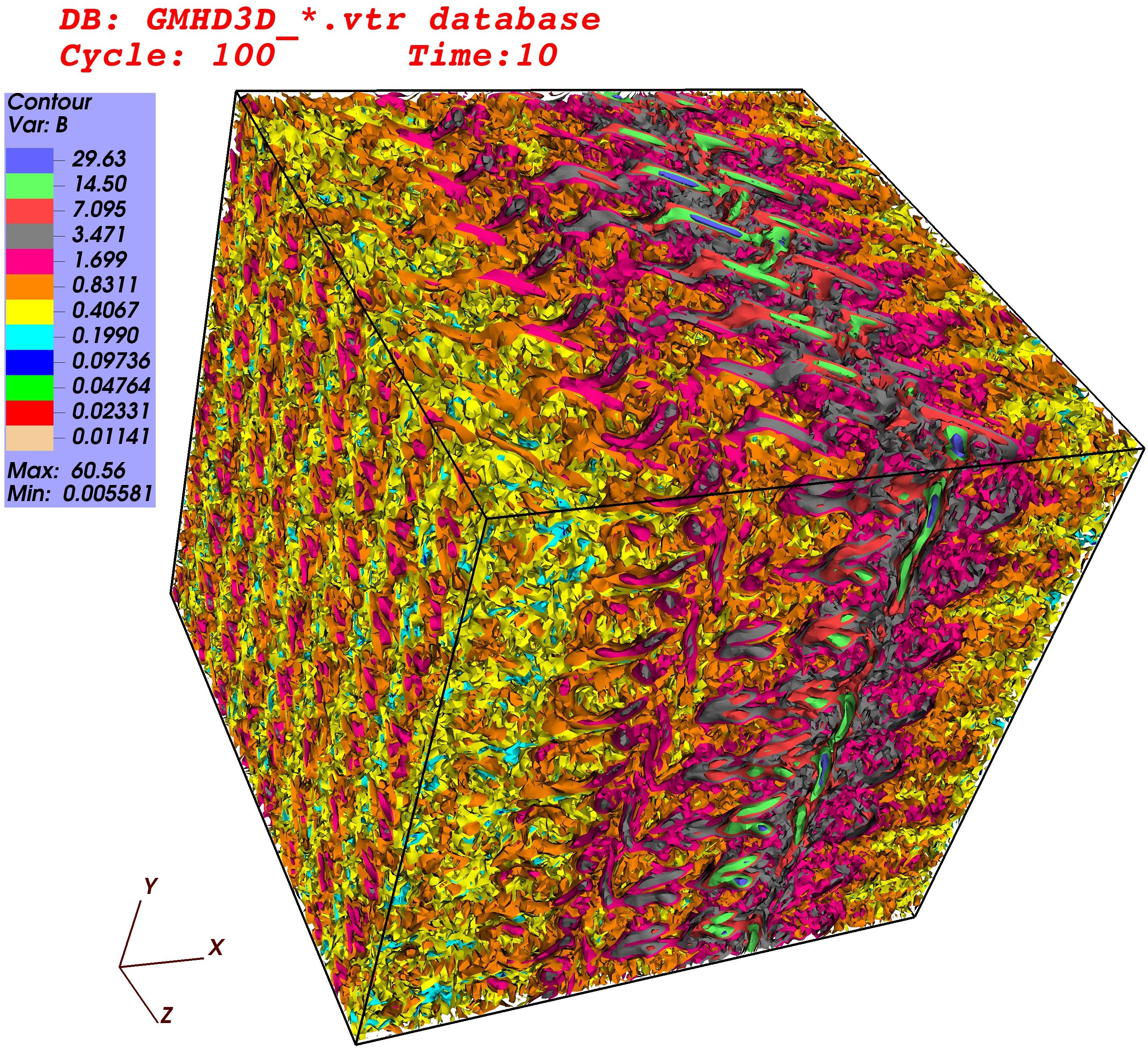}
		\caption{Time = 10.0}
		%\label{initial flow beta 0p4}
	\end{subfigure}
	\begin{subfigure}{0.49\textwidth}
		\centering
		\includegraphics[scale=0.08480]{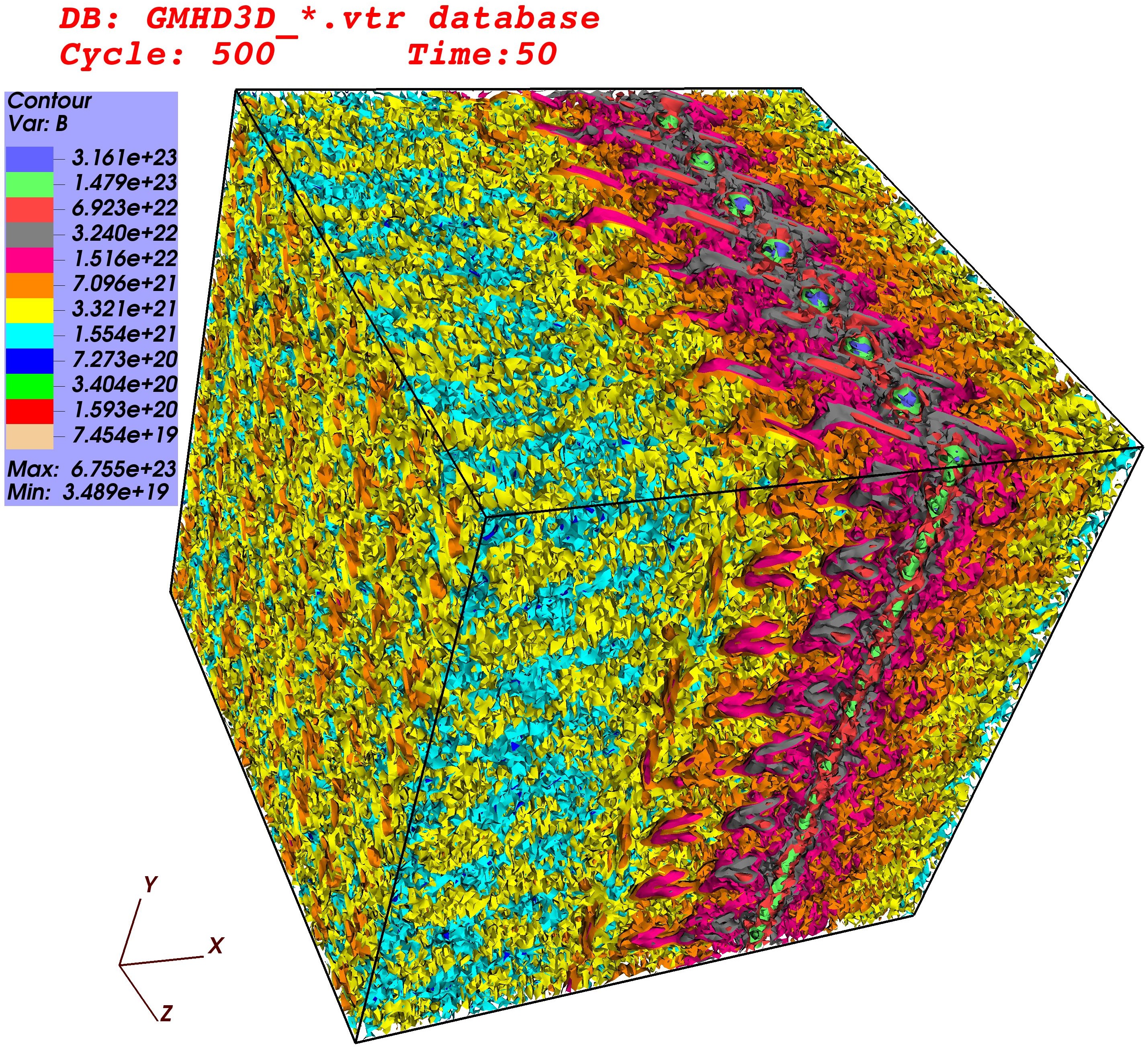}
		\caption{Time = 50.0}
		%\label{initial flow beta 0p4}
	\end{subfigure}
	\caption{Time evolution of 3-dimensional magnetic field iso-surfaces (Iso-B surfaces) for a given small scale EPI2D flow ($k_0 = 8$) and  broken jet flow shear ($k_s \to \infty$. i.e, $k_s \sim k_{max}$ numerically speaking). The structures are mostly dominated by short scale structures (SSD) [\textbf{Movie3.mp4}]. Simulation details: grid resolution $256^3$, stepping time $dt = 10^{-4}$, shear flow strength $S = 5.0$, magnetic Reynolds number $R_m = 200.0$. Visualization in log scale.}
		\label{Strip Shear magnetic field evolution}
\end{figure*}

%Strip Vortex
A natural question is whether the dynamo property found here is associated to the scale of the shear flow profile. To answer this question, we have conducted numerical experiments using broken jet flow shear as an extreme case (i.e, $k_s \to \infty$ or, $k_s \sim k_{max}$ numerically speaking) (See Fig. \ref{initial ABC flow}d). This shear profile appears frequently in hydrodynamics studies, especially those used to investigate Navier-Stokes turbulence \cite{Drazin:1961, Thess_Strip:1994, Shishir_POF:2022}. EPI2D flow is shown to generate magnetic energy ($E_B = \frac{1}{2} \int_{V} (B_x^2 + B_y^2 + B_z^2) dx dy dz$) in an exponential fashion in the presence of broken jet flow (with $k_s \sim k_{max}$)  shear. Furthermore, from Fig. \ref{EPI2D Strip type}a, it is clear that with increase in magnetic Reynolds number $R_m$, the growth rate of magnetic energy ($\gamma = \frac{d}{dt}(\ln E_B(t))$) is found not to reach a saturation point in line with what is expected for a fast dynamo action. With the help of an iso-surface representation of magnetic energy (See Fig.\ref{Strip Shear magnetic field evolution}), we are able to determine that the generated magnetic energy is mostly contained in smaller scales (compared to the length scale of the flow that is driving it), making the dynamo a small-scale dynamo (SSD). In addition, we have estimated the magnetic energy spectral density associated with each mode, such as $|\hat{B}(k = 1), t|, |\hat{B}(k = 20), t|, |\hat{B}(k = 30), t|$ etc., and monitor its time-dependent evolution. As was the case in the previous example, it is found that the energy is  concentrated in higher modes or shorter length scales (See Fig. \ref{EPI2D Strip type}b); consequently, the dynamo can be thought of as a small-scale dynamo (SSD).
We also plot the magnetic energy growth rate ($\gamma = \frac{d}{dt}(\ln E_B(t))$) as a function of shear flow strength ($S$), and we obtain the same generalized algebraic scaling $\gamma(S) = - aS + bS^\frac{2}{3}$ \textcolor{black}{(based on $\chi^2$ minimization)} , where $a$ \& $b$ are real fit coefficients as obtained for broken jet shear flow (See Fig. \ref{EPI2D Strip type}c). Hence, in the presence of broken jet flow shear \cite{Drazin:1961, Thess_Strip:1994, Shishir_POF:2022} as well, our numerical observation unambiguously demonstrates that there is an onset and increase in the activity of small scale dynamo (SSD). As the effect is found to be robust at largest and smallest shear scale lengths, we conclude that the scaling of $\gamma(S)$ vs $S$ appears to be independent of shear flow scale (except that the real coefficients a and b are to be determined accordingly) and robust. 

To further convince ourselves regarding the generality of our finding, we have considered yet another non-helical flow namely Taylor-Green (TG) flow and investigate the effect of flow shear on the dynamo activity with TG flow as the base flow.  It is found that the fundamental observations remain unaltered (See Supplementary information for details).

\section{Summary and Conclusion}
In this work, we have performed direct numerical simulations (DNS) of kinematic dynamos using a 3-dimensional magnetohydrodynamic model at \textcolor{black}{modest} grid resolutions. By considering a simple kinematic dynamo model, we are able to demonstrate that the small-scale helical YM flow with $\beta = 1.0$ (ABC-like flow) generates rapid dynamo action over a broad range of magnetic Reynolds number. The results of our spectral calculation demonstrate without a doubt that the fully developed dynamo is, in effect, a \textcolor{black}{small} scale dynamo (SSD). In addition, it has been shown that the presence of a  flow shear reduces the efficiency of small-scale dynamo action. This is interesting finding appears to be consistent with several earlier works reported for a 2.5-dimensional GP flow, which may also be considered as a good benchmark for GMHD3D solver.\\

Our major findings are: \\

$\bullet$ A non-dynamo producing, small scale non-helical EPI2D flow shows fast SSD activity when flow shear is introduced. More importantly, unlike fully helical flows, it has been observed that, for EPI2D non-helical flows, the small-scale dynamo action (SSD) increases as the shear flow strength ($S$) increases.  The spectral diagnostics also are found to be in agreement with the observation. %Compared to earlier findings, we believe that this is a nontrivial and intriguing finding.

%By employing a kinematic dynamo model, we have identified that flow shear can enhance the efficiency of small-scale dynamos for a given non-helical velocity field.

$\bullet$ A generalized algebraic scaling for the magnetic energy growth rate ($\gamma$) as a function of the shear flow strength ($S$) has been obtained. Our numerical observation is supported by a number of recent analytical works \cite{Kolokolov:2011,Proctor_JFM:2012}.

%$\bullet$ \textcolor{red}{Using a standard small-scale base flow, we have efficiently explained from our numerical experiment that, when fluid helicity is injected into the system, flow shear is effectively trying to suppress small scale dynamo (SSD) activity.}

$\bullet$ We have performed our numerical experiments using broken jet flow ($k_s \to \infty$. i.e, $k_s \sim k_{max}$ numerically speaking) shear profile, and we have found that the primary findings are unaffected. The mechanism of onset of dynamo from non-helical base flows in the presence of shear flows is found to be independent of the scale of shear flows. 

$\bullet$ The scaling of $\gamma (S) = - aS + bS^\frac{2}{3}$, where $a$ \& $b$ are real fit coefficients is found be to robust and not dependent on shear flow length scale $k_s$. 

$\bullet$ We have also carried out the same analysis for a different well-known non-helical base flow, known as the Taylor-Green flow, and found that our results remain valid (See Supplementary information for details). Our numerical findings are basically found to be robust. 
%(which in the absence of flow shear is a non-dynamo flow)

To conclude, we have investigated the effect of shear flows on the onset of dynamo instability on non-helical base flows, using a \textcolor{black}{modest} resolution and a wide range of magnetic Reynolds numbers and for flow shear scale lengths. Our numerical analysis reveals that the small-scale dynamo action is suppressed by flow shear for base helical flows, but is amplified for base non-helical flows. We also believe these dynamos should play an important role in a wide variety of astrophysical objects, especially in highly symmetric regions like the mid plane of accretion disks where flow shear is  known to be substantial \cite{MRI_Balbus:1998, MRI_Donati:2005}. \textcolor{black}{However, it may be argued that in actual astrophysical conditions, the magnetic back reaction on the velocity field cannot be disregarded. Unlike the scenario presented above, velocity fields in such situations are not predetermined. We hope to address several of these issues in a future communication.}

%=========================================================================================

%=========================================================================================
\section{ACKNOWLEDGMENTS}
The simulations and visualizations presented here are performed on GPU nodes and visualization nodes of Antya cluster at the Institute for Plasma Research (IPR), INDIA. One of the author S.B is thankful to Dr. Rupak Mukherjee at Central University of Sikkim (CUS), Gangtok, Sikkim, India for providing an initial version of \textit{GMHD3D} code. S.B thanks N. Vydyanathan, Bengaluru and B. K. Sharma at NVIDIA, Bengaluru, India, for extending their help with  basic GPU methods. S.B is grateful to Mr. Soumen De Karmakar at IPR, India for many helpful discussions regarding GPUs, and HPC support team of IPR for extending their help related to ANTYA cluster. %The authors would also like to thank Dr. Jugal Chowdhury from IPR for careful reviewing the manuscript and for his valuable suggestions.
%=========================================================================================

\section{\textbf{DATA AVAILABILITY}}
The data underlying this article will be shared on reasonable request to the corresponding author.

\section{\textbf{Conflict of Interest}}
The authors have no conflicts to disclose.

\section{\textbf{SUPPORTING INFORMATION}}
The following movies:
\begin{itemize}
	%	\item \textbf{Multimedia\_3b.mp4}
	\item \textbf{Movie1.mp4}
	\item \textbf{Movie2.mp4}
	\item \textbf{Movie3.mp4}
	%	\item \textbf{Multimedia\_6b.mp4}
	%	\item \textbf{Multimedia\_7b.mp4}
\end{itemize}
are added as integral multimedia files.

%\section{References}
\bibliography{references}

\onecolumngrid
\appendix
\section{\textbf{Benchmarking of GMHD3D solver}}\label{Appen A}
\begin{figure*}[h]
	\centering
	\begin{subfigure}{0.49\textwidth}
		\centering
		%\hfill
		\includegraphics[scale=0.6]{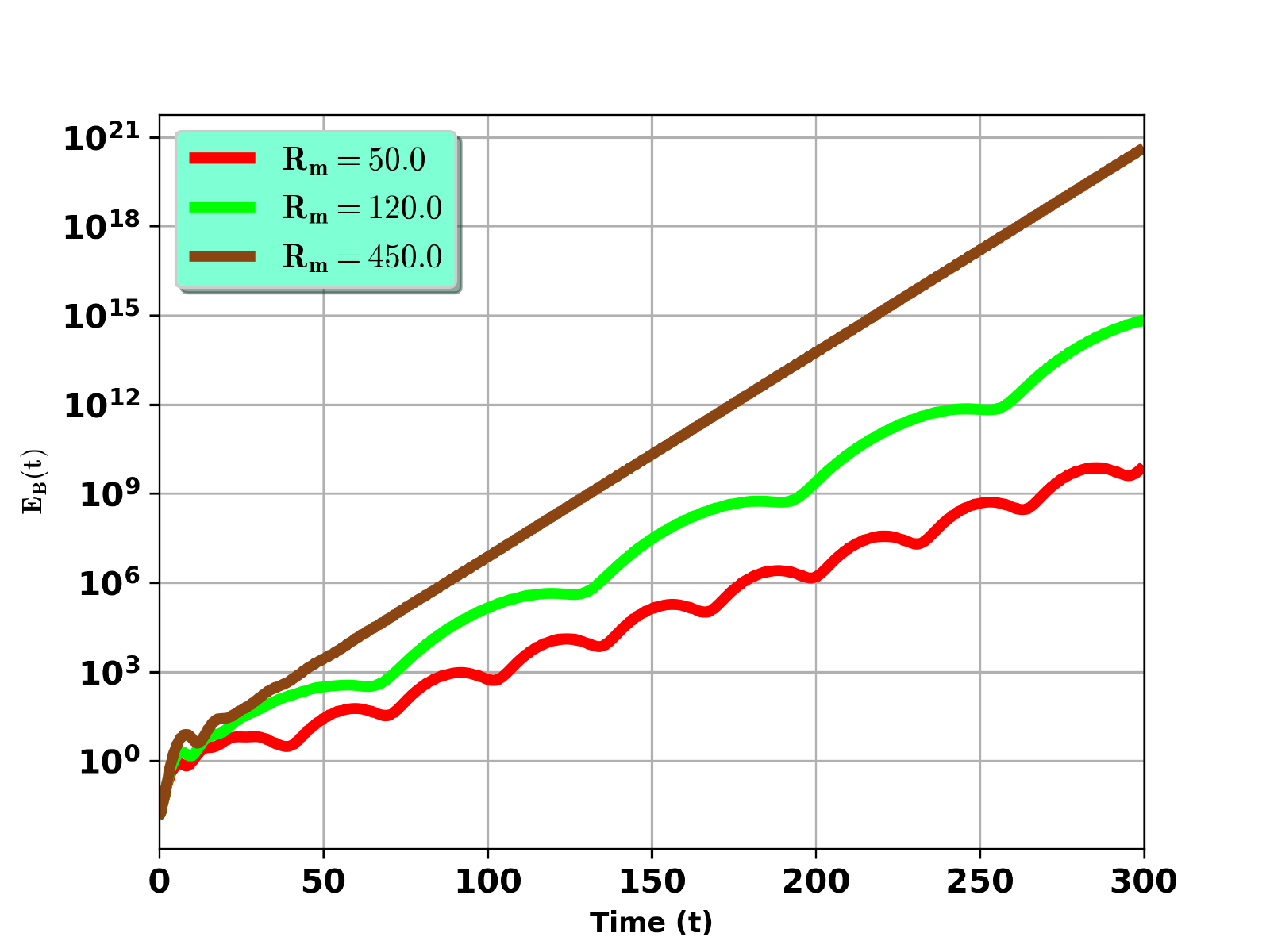}
		\caption{}
		%	\label{initial flow beta 0}
	\end{subfigure}
	\begin{subfigure}{0.49\textwidth}
		\centering
		\includegraphics[scale=0.07]{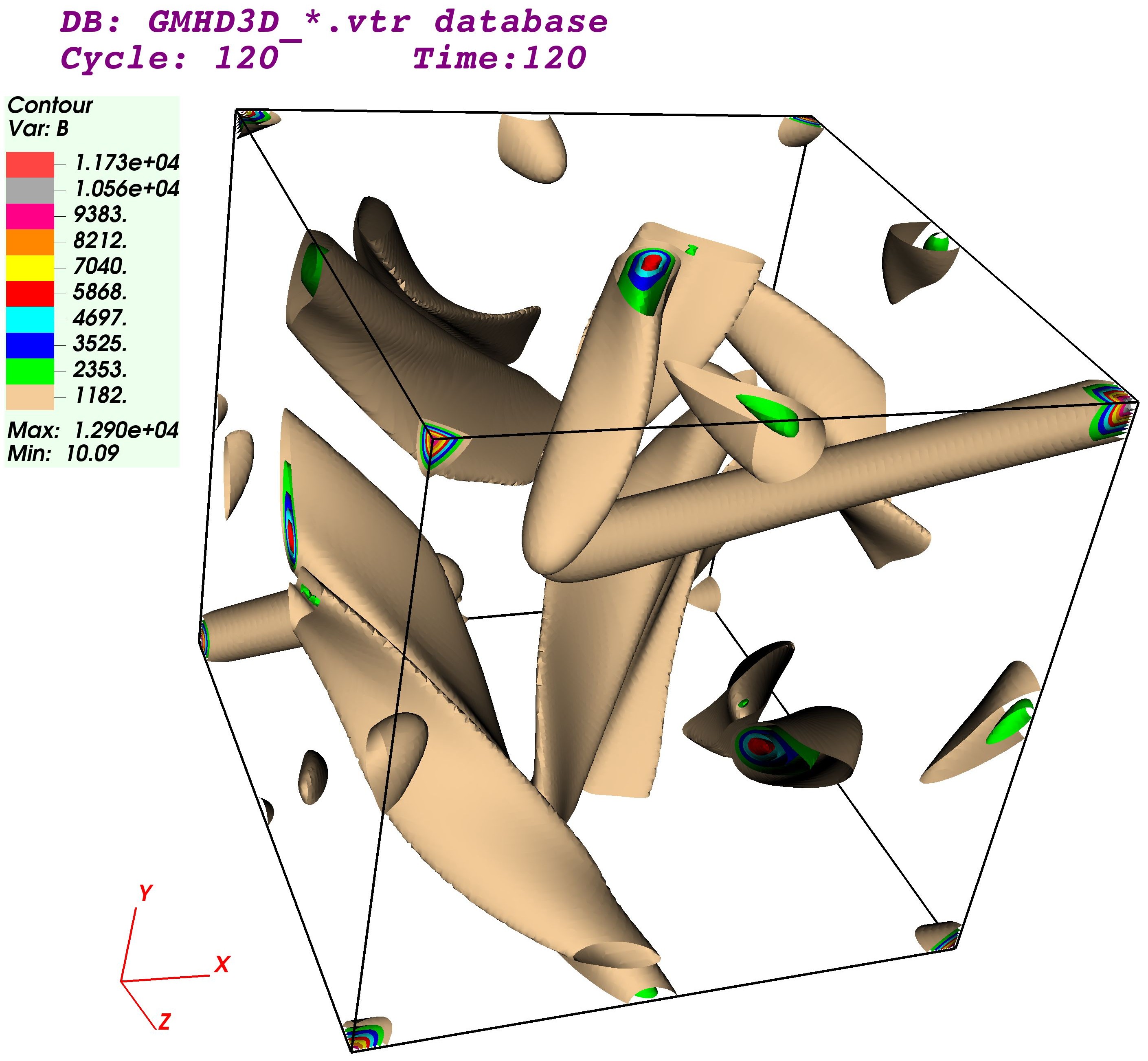}
		\caption{}
		%	\label{initial flow beta 0p2}
	\end{subfigure}
	\caption{\textcolor{black}{(a) Time evolution of magnetic energy ($E_B$) for different values of magnetic Reynolds numbers. (b) Magnetic energy iso-surface visualization leads to concentrated ``cigar like'' structures. The identical observation was reported by \citet{Frish_Dynamo:1986}.}}
	\label{ABC111}
\end{figure*}

\textcolor{black}{We have looked into the kinematic dynamo problem for a class of ABC flow at different magnetic Reynolds numbers. The velocity profile for ABC flow is given by,
	\begin{equation}\label{ABC}
		\begin{aligned}
			u_x &=  [ A \sin(y) -  C \cos(z) ]\\
			u_y &=  [  B \sin(z) -  A \cos(x) ]\\
			u_z &=  [  C \sin(x) -  B \cos(y) ]
		\end{aligned}
	\end{equation}
	where the choice of $A, B, C$ values are $1$. When we use this initial condition, we obtain the identical exponential growth in magnetic energy like \citet{Frish_Dynamo:1986}. The three-dimensional visualization of the magnetic energy iso-surface reveals ``cigar-like'' structures, which represent the fastest-growing growth mode [See Fig. \ref{ABC111}].
}

\begin{figure*}[h]
	\centering
	\begin{subfigure}{0.49\textwidth}
		\centering
		%\hfill
		\includegraphics[scale=0.6]{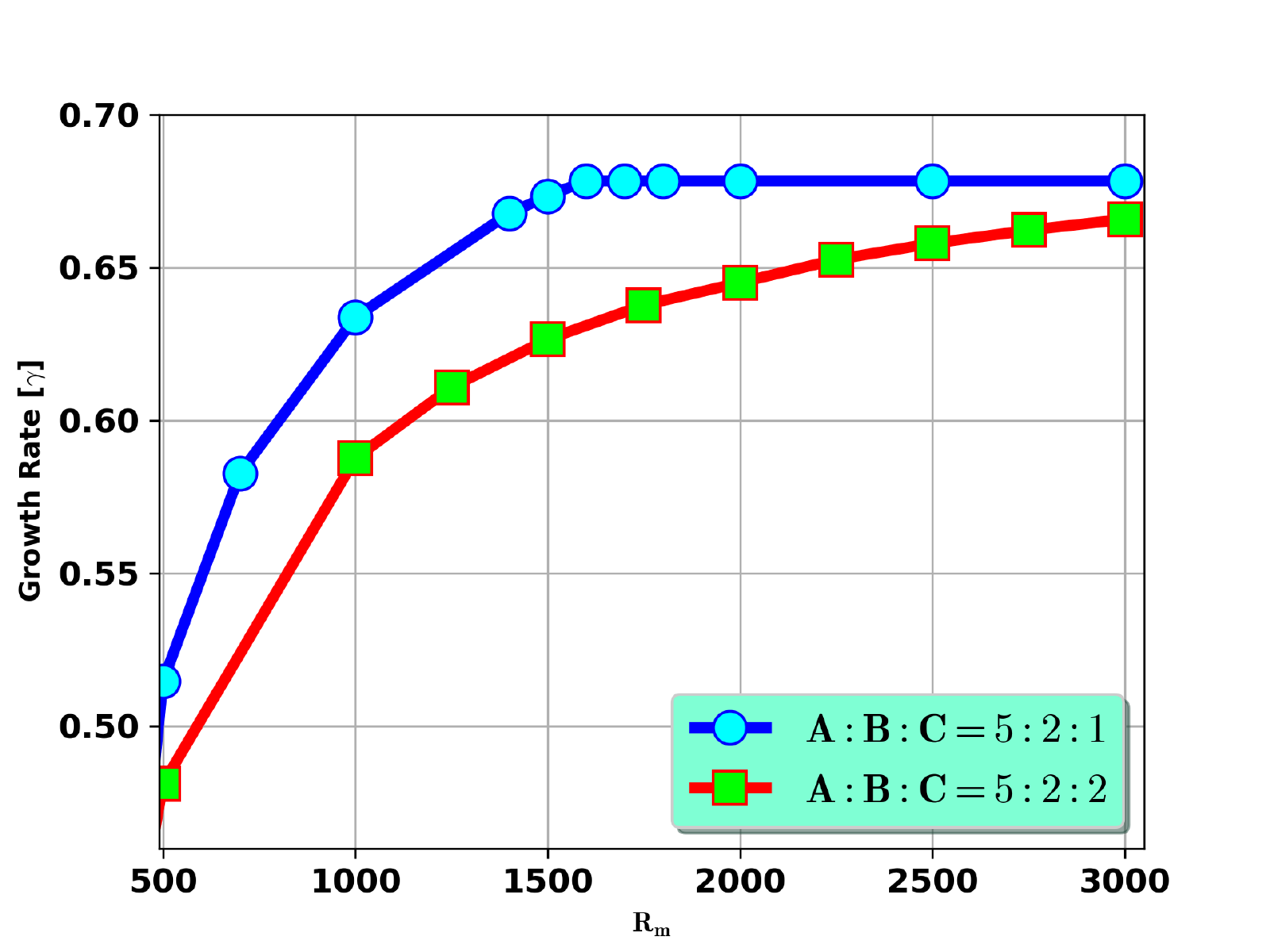}
		\caption{}
		%	\label{initial flow beta 0}
	\end{subfigure}
	\begin{subfigure}{0.49\textwidth}
		\centering
		\includegraphics[scale=0.07]{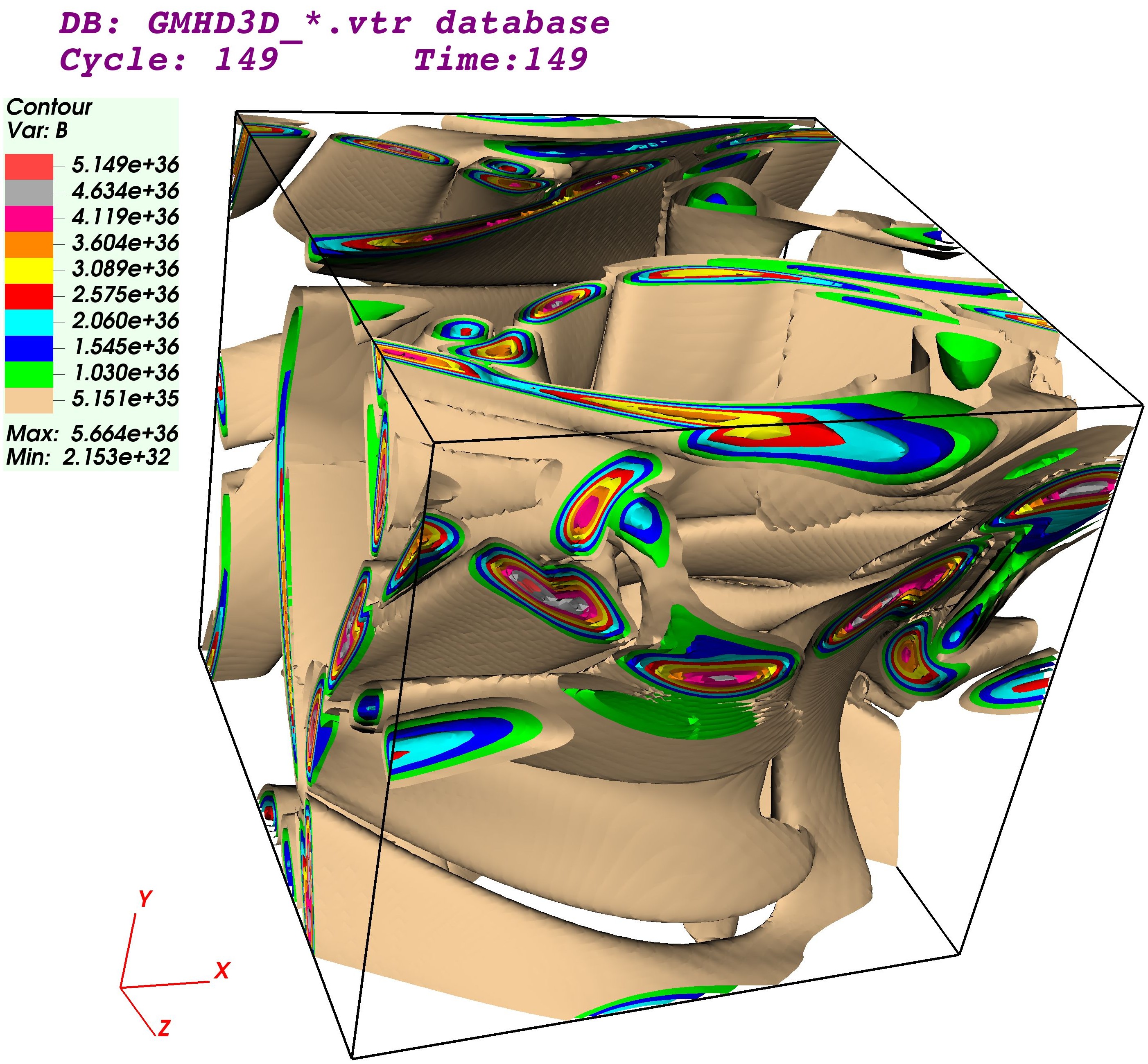}
		\caption{}
		%	\label{initial flow beta 0p2}
	\end{subfigure}
	\caption{\textcolor{black}{(a) Magnetic energy growth rate ($\gamma$) versus the magnetic
			Reynolds number ($R_m$) for the $5 : 2 : 1$ \& $5 : 2 : 2$ flow. (b) The magnetic energy iso-surface looks like ribbons when $A : B : C = 5 : 2 : 2$. This observation is identical to Archontis et al \cite{Archontis:2003}.}}
	\label{ABC522}
\end{figure*}

\textcolor{black}{We have also used ABC flow to study the properties of kinematic dynamos, but with a slightly distinct topology. We have investigated the case where $A, B, C$ is $5, 2, 2$ and found that in the latter instance, the magnetic field is amplified exponentially, with the growth rate increasing with $R_m$. However, in the $5,2,1$ scenario, it is discovered that the growth rate of magnetic energy reaches a maximum at a value of $0.67$. Archontis et al \cite{Archontis:2003} also found something very close to this. The ``cigar-like'' structures that form in the classical $A = B = C =1$ case are substituted with ``ribbon-like'' structures [See Fig. \ref{ABC522}].
}

\end{document}